\input harvmac
\let\includefigures=\iftrue
\let\useblackboard==\iftrue
\newfam\black

\includefigures

\let\includefigures=\iftrue
\let\useblackboard=\iftrue
\newfam\black

\includefigures
\message{If you do not have epsf.tex (to include figures),}
\message{change the option at the top of the tex file.}
\input epsf
\def\figin{\epsfcheck\figin}\def\figins{\epsfcheck\figins}
\def\epsfcheck{\ifx\epsfbox\UnDeFiNeD
\message{(NO epsf.tex, FIGURES WILL BE IGNORED)}
\gdef\figin##1{\vskip2in}\gdef\figins##1{\hskip.5in}
\else\message{(FIGURES WILL BE INCLUDED)}%
\gdef\figin##1{##1}\gdef\figins##1{##1}\fi}
\def\DefWarn#1{}
\def\figinsert{\goodbreak\midinsert}
\def\ifig#1#2#3{\DefWarn#1\xdef#1{fig.~\the\figno}
\writedef{#1\leftbracket fig.\noexpand~\the\figno}%
\figinsert\figin{\centerline{#3}}\medskip\centerline{\vbox{
\baselineskip12pt\advance\hsize by -1truein
\noindent\footnotefont{\bf Fig.~\the\figno:} #2}}
\endinsert\global\advance\figno by1}
\else
\def\ifig#1#2#3{\xdef#1{fig.~\the\figno}
\writedef{#1\leftbracket fig.\noexpand~\the\figno}%
\global\advance\figno by1} \fi

\def\journal#1&#2(#3){\unskip, \sl #1\ \bf #2 \rm(19#3) }
\def\andjournal#1&#2(#3){\sl #1~\bf #2 \rm (19#3) }

\noblackbox
%


\def\unlockat{\catcode`\@=11}
\def\lockat{\catcode`\@=12}

\unlockat

\def\newsec#1{\global\advance\secno by1\message{(\the\secno. #1)}
\global\subsecno=0\global\subsubsecno=0\eqnres@t\noindent
{\bf\the\secno. #1}
\writetoca{{\secsym} {#1}}\par\nobreak\medskip\nobreak}
\global\newcount\subsecno \global\subsecno=0
\def\subsec#1{\global\advance\subsecno
by1\message{(\secsym\the\subsecno. #1)}
\ifnum\lastpenalty>9000\else\bigbreak\fi\global\subsubsecno=0
\noindent{\it\secsym\the\subsecno. #1}
\writetoca{\string\quad {\secsym\the\subsecno.} {#1}}
\par\nobreak\medskip\nobreak}
\global\newcount\subsubsecno \global\subsubsecno=0
\def\subsubsec#1{\global\advance\subsubsecno by1
\message{(\secsym\the\subsecno.\the\subsubsecno. #1)}
\ifnum\lastpenalty>9000\else\bigbreak\fi
\noindent\quad{\secsym\the\subsecno.\the\subsubsecno.}{#1}
\writetoca{\string\qquad{\secsym\the\subsecno.\the\subsubsecno.}{#1}}
\par\nobreak\medskip\nobreak}

\def\subsubseclab#1{\DefWarn#1\xdef
#1{\noexpand\hyperref{}{subsubsection}%
{\secsym\the\subsecno.\the\subsubsecno}%
{\secsym\the\subsecno.\the\subsubsecno}}%
\writedef{#1\leftbracket#1}\wrlabeL{#1=#1}}
\lockat

%


\font\manual=manfnt \def\dbend{\lower3.5pt\hbox{\manual\char127}}

\def\IZ{\relax\ifmmode\mathchoice
{\hbox{\cmss Z\kern-.4em Z}}{\hbox{\cmss Z\kern-.4em Z}}
{\lower.9pt\hbox{\cmsss Z\kern-.4em Z}}
{\lower1.2pt\hbox{\cmsss Z\kern-.4em Z}}\else{\cmss Z\kern-.4em
Z}\fi}
\def\half{{1\over 2}}


\def\IZ{\relax\ifmmode\mathchoice
{\hbox{\cmss Z\kern-.4em Z}}{\hbox{\cmss Z\kern-.4em Z}}
{\lower.9pt\hbox{\cmsss Z\kern-.4em Z}}
{\lower1.2pt\hbox{\cmsss Z\kern-.4em Z}}\else{\cmss Z\kern-.4em
Z}\fi}
\def\IB{\relax{\rm I\kern-.18em B}}
\def\IC{{\relax\hbox{$\inbar\kern-.3em{\rm C}$}}}
\def\ID{\relax{\rm I\kern-.18em D}}
\def\IE{\relax{\rm I\kern-.18em E}}
\def\IF{\relax{\rm I\kern-.18em F}}
\def\IG{\relax\hbox{$\inbar\kern-.3em{\rm G}$}}
\def\IGa{\relax\hbox{${\rm I}\kern-.18em\Gamma$}}
\def\IH{\relax{\rm I\kern-.18em H}}
\def\II{\relax{\rm I\kern-.18em I}}
\def\IK{\relax{\rm I\kern-.18em K}}
\def\IP{\relax{\rm I\kern-.18em P}}
\def\IQ{\relax\hbox{$\inbar\kern-.3em{\rm Q}$}}

\def\inbar{\,\vrule height1.5ex width.4pt depth0pt}

\font\cmss=cmss10 \font\cmsss=cmss10 at 7pt
\def\IR{\relax{\rm I\kern-.18em R}}

%
%

\def\makeblankbox#1#2{\hbox{\lower\dp0\vbox{\hidehrule{#1}{#2}%
   \kern -#1
   \hbox to \wd0{\hidevrule{#1}{#2}%
      \raise\ht0\vbox to #1{}
      \lower\dp0\vtop to #1{}
      \hfil\hidevrule{#2}{#1}}%
   \kern-#1\hidehrule{#2}{#1}}}%
}%
\def\hidehrule#1#2{\kern-#1\hrule height#1 depth#2 \kern-#2}%
\def\hidevrule#1#2{\kern-#1{\dimen0=#1\advance\dimen0 by #2\vrule
    width\dimen0}\kern-#2}%
\def\openbox{\ht0=1.2mm \dp0=1.2mm \wd0=2.4mm  \raise 2.75pt
\makeblankbox {.25pt} {.25pt}  }

\def\bun#1/#2{\leavevmode
   \kern.1em \raise .5ex \hbox{\the\scriptfont0 #1}%
   \kern-.1em $/$%
   \kern-.15em \lower .25ex \hbox{\the\scriptfont0 #2}%
}

\def\opensquare{\ht0=3.4mm \dp0=3.4mm \wd0=6.8mm  \raise 2.7pt
\makeblankbox {.25pt} {.25pt}  }


\def\sector#1#2{\ {\scriptstyle #1}\hskip 1mm
\mathop{\opensquare}\limits_{\lower 1mm\hbox{$\scriptstyle#2$}}\hskip 1mm}

\def\tsector#1#2{\ {\scriptstyle #1}\hskip 1mm
\mathop{\opensquare}\limits_{\lower 1mm\hbox{$\scriptstyle#2$}}^\sim\hskip 1mm}


\def\inbar{\,\vrule height1.5ex width.4pt depth0pt}

\font\cmss=cmss10 \font\cmsss=cmss10 at 7pt
\def\IR{\relax{\rm I\kern-.18em R}}


\def\frac#1#2{{#1\over#2}}

\def\half{\frac12}

\def\inbar{\,\vrule height1.5ex width.4pt depth0pt}
\def\IC{\relax\hbox{$\inbar\kern-.3em{\rm C}$}}
\def\IR{\relax{\rm I\kern-.18em R}}
\def\IP{\relax{\rm I\kern-.18em P}}

%
%
\catcode`\@=11
\def\slash#1{\mathord{\mathpalette\c@ncel{#1}}}
\overfullrule=0pt

\def\II{{\cal I}}

\def\LL{{\cal L}}

\def\NN{{\cal N}}

\def\underrel#1\over#2{\mathrel{\mathop{\kern\z@#1}\limits_{#2}}}

\catcode`\@=12


%

\def\sinh{{\rm sinh}}
\def\cosh{{\rm cosh}}

\def\exp{{\rm exp}}



\def\frac#1#2{{#1\over#2}}

\def\half{\frac12}

\def\inbar{\,\vrule height1.5ex width.4pt depth0pt}
\def\IC{\relax\hbox{$\inbar\kern-.3em{\rm C}$}}
\def\IR{\relax{\rm I\kern-.18em R}}
\def\IP{\relax{\rm I\kern-.18em P}}

%
%

%
\catcode`\@=11
\def\slash#1{\mathord{\mathpalette\c@ncel{#1}}}
\overfullrule=0pt

\def\II{{\cal I}}

\def\LL{{\cal L}}

\def\NN{{\cal N}}

\def\underrel#1\over#2{\mathrel{\mathop{\kern\z@#1}\limits_{#2}}}

\catcode`\@=12


%

\def \sinh{{\rm sinh}}
\def \cosh{{\rm cosh}}

\def\exp{{\rm exp}}


\lref\WakimotoGF{
  M.~Wakimoto,
  ``Fock representations of the affine lie algebra A1(1),''
Commun.\ Math.\ Phys.\  {\bf 104}, 605 (1986).
}

\lref\BarsRB{
  I.~Bars and D.~Nemeschansky,
  ``String Propagation in Backgrounds With Curved Space-time,''
Nucl.\ Phys.\ B {\bf 348}, 89 (1991).
}

\lref\KazakovPM{
  V.~Kazakov, I.~K.~Kostov and D.~Kutasov,
  ``A Matrix model for the two-dimensional black hole,''
Nucl.\ Phys.\ B {\bf 622}, 141 (2002).
[hep-th/0101011].
}
\lref\BershadskyMF{
  M.~Bershadsky and H.~Ooguri,
  ``Hidden SL(n) Symmetry in Conformal Field Theories,''
Commun.\ Math.\ Phys.\  {\bf 126}, 49 (1989).
}

\lref\MathurJK{
  S.~D.~Mathur and D.~Turton,
  ``Comments on black holes I: The possibility of complementarity,''
JHEP {\bf 1401}, 034 (2014).
[arXiv:1208.2005 [hep-th]].
}

\lref\AveryTF{
  S.~G.~Avery, B.~D.~Chowdhury and A.~Puhm,
  ``Unitarity and fuzzball complementarity: 'Alice fuzzes but may not even know it!',''
JHEP {\bf 1309}, 012 (2013).
[arXiv:1210.6996 [hep-th]].
}

\lref\AharonyXN{
  O.~Aharony, A.~Giveon and D.~Kutasov,
  ``LSZ in LST,''
Nucl.\ Phys.\ B {\bf 691}, 3 (2004).
[hep-th/0404016].
}

\lref\PolchinskiTA{
  J.~Polchinski,
  ``String theory and black hole complementarity,''
[hep-th/9507094].
}

\lref\GerasimovFI{
  A.~Gerasimov, A.~Morozov, M.~Olshanetsky, A.~Marshakov and S.~L.~Shatashvili,
  ``Wess-Zumino-Witten model as a theory of free fields,''
Int.\ J.\ Mod.\ Phys.\ A {\bf 5}, 2495 (1990).
}

\lref\ItzhakiJT{
  N.~Itzhaki,
  ``Is the black hole complementarity principle really necessary?,''
[hep-th/9607028].
}

\lref\BraunsteinMY{
  S.~L.~Braunstein, S.~Pirandola and K.~Życzkowski,
  ``Better Late than Never: Information Retrieval from Black Holes,''
Phys.\ Rev.\ Lett.\  {\bf 110}, no. 10, 101301 (2013).
[arXiv:0907.1190 [quant-ph]].
}

\lref\WittenZW{
  E.~Witten,
  ``Anti-de Sitter space, thermal phase transition, and confinement in gauge theories,''
Adv.\ Theor.\ Math.\ Phys.\  {\bf 2}, 505 (1998).
[hep-th/9803131].
}

\lref\SusskindUW{
  L.~Susskind,
  ``The Transfer of Entanglement: The Case for Firewalls,''
[arXiv:1210.2098 [hep-th]].
}

\lref\PapadodimasAQ{
  K.~Papadodimas and S.~Raju,
  ``An Infalling Observer in AdS/CFT,''
JHEP {\bf 1310}, 212 (2013).
[arXiv:1211.6767 [hep-th]].
}
\lref\GoulianQR{
  M.~Goulian and M.~Li,
  ``Correlation functions in Liouville theory,''
Phys.\ Rev.\ Lett.\  {\bf 66}, 2051 (1991).
}

\lref\AlmheiriRT{
  A.~Almheiri, D.~Marolf, J.~Polchinski and J.~Sully,
  ``Black Holes: Complementarity or Firewalls?,''
JHEP {\bf 1302}, 062 (2013).
[arXiv:1207.3123 [hep-th]].
}

\lref\MarolfDBA{
  D.~Marolf and J.~Polchinski,
  ``Gauge/Gravity Duality and the Black Hole Interior,''
Phys.\ Rev.\ Lett.\  {\bf 111}, 171301 (2013).
[arXiv:1307.4706 [hep-th]].
}

\lref\PolchinskiCEA{
  J.~Polchinski,
  ``Chaos in the black hole S-matrix,''
[arXiv:1505.08108 [hep-th]].
}

\lref\GiveonGFK{
  A.~Giveon and N.~Itzhaki,
  ``Stringy Black Hole Interiors,''
JHEP {\bf 1911}, 014 (2019).
[arXiv:1908.05000 [hep-th]].
}

\lref\MathurHF{
  S.~D.~Mathur,
  ``The Information paradox: A Pedagogical introduction,''
Class.\ Quant.\ Grav.\  {\bf 26}, 224001 (2009).
[arXiv:0909.1038 [hep-th]].
}

\lref\ItzhakiTU{
  N.~Itzhaki, D.~Kutasov and N.~Seiberg,
  ``I-brane dynamics,''
JHEP {\bf 0601}, 119 (2006).
[hep-th/0508025].
}

\lref\DvaliRT{
  G.~Dvali and C.~Gomez,
  ``Black Hole's 1/N Hair,''
Phys.\ Lett.\ B {\bf 719}, 419 (2013).
[arXiv:1203.6575 [hep-th]].
}

\lref\DvaliAA{
  G.~Dvali and C.~Gomez,
  ``Black Hole's Quantum N-Portrait,''
Fortsch.\ Phys.\  {\bf 61}, 742 (2013).
[arXiv:1112.3359 [hep-th]].
}

\lref\BershadskyIN{
  M.~Bershadsky and D.~Kutasov,
  ``Comment on gauged WZW theory,''
Phys.\ Lett.\ B {\bf 266}, 345 (1991).
}

\lref\GiveonCMA{
  A.~Giveon, N.~Itzhaki and D.~Kutasov,
  ``Stringy Horizons,''
JHEP {\bf 1506}, 064 (2015).
[arXiv:1502.03633 [hep-th]].
}

\lref\HartleAI{
  J.~B.~Hartle and S.~W.~Hawking,
  ``Wave Function of the Universe,''
Phys.\ Rev.\ D {\bf 28}, 2960 (1983), [Adv.\ Ser.\ Astrophys.\ Cosmol.\  {\bf 3}, 174 (1987)].
}

\lref\GrossKZA{
  D.~J.~Gross and P.~F.~Mende,
  ``The High-Energy Behavior of String Scattering Amplitudes,''
Phys.\ Lett.\ B {\bf 197}, 129 (1987).
}
\lref\GrossAR{
  D.~J.~Gross and P.~F.~Mende,
  ``String Theory Beyond the Planck Scale,''
Nucl.\ Phys.\ B {\bf 303}, 407 (1988).
}

\lref\ItzhakiGLF{
  N.~Itzhaki,
  ``Stringy instability inside the black hole,''
JHEP {\bf 1810}, 145 (2018).
[arXiv:1808.02259 [hep-th]].
}

\lref\ItzhakiRLD{
  N.~Itzhaki and L.~Liram,
  ``A stringy glimpse into the black hole horizon,''
JHEP {\bf 1804}, 018 (2018).
[arXiv:1801.04939 [hep-th]].
}

\lref\Ben{
  R.~Ben , A.~Giveon, N.~Itzhaki and L.~Liram,
  ``On the black hole interior in string theory,''
JHEP {\bf 1705}, 094 (2017).
[arXiv:1702.03583 [hep-th]].
}

\lref\PolyakovVU{
  A.~M.~Polyakov,
  ``Thermal Properties of Gauge Fields and Quark Liberation,''
Phys.\ Lett.\  {\bf 72B}, 477 (1978).
}

\lref\SusskindUP{
  L.~Susskind,
  ``Lattice Models of Quark Confinement at High Temperature,''
Phys.\ Rev.\ D {\bf 20}, 2610 (1979).
}

\lref\MaldacenaXJA{
  J.~Maldacena and L.~Susskind,
  ``Cool horizons for entangled black holes,''
Fortsch.\ Phys.\  {\bf 61}, 781 (2013).
[arXiv:1306.0533 [hep-th]].
}

\lref\KutasovXU{
  D.~Kutasov and N.~Seiberg,
  ``More comments on string theory on AdS(3),''
JHEP {\bf 9904}, 008 (1999).
[hep-th/9903219].
}

\lref\TeschnerUG{
  J.~Teschner,
  ``Operator product expansion and factorization in the H+(3) WZNW model,''
Nucl.\ Phys.\ B {\bf 571}, 555 (2000).
[hep-th/9906215].
}

\lref\GiribetFT{
  G.~Giribet and C.~A.~Nunez,
  ``Correlators in AdS(3) string theory,''
JHEP {\bf 0106}, 010 (2001).
[hep-th/0105200].
}

\lref\HorowitzNW{
  G.~T.~Horowitz and J.~Polchinski,
  ``A Correspondence principle for black holes and strings,''
Phys.\ Rev.\ D {\bf 55}, 6189 (1997).
[hep-th/9612146].
}

\lref\GiveonHFA{
  A.~Giveon, N.~Itzhaki and J.~Troost,
  ``Lessons on Black Holes from the Elliptic Genus,''
JHEP {\bf 1404}, 160 (2014).
[arXiv:1401.3104 [hep-th]].
}

\lref\SusskindWS{
  L.~Susskind,
  ``Some speculations about black hole entropy in string theory,''
In *Teitelboim, C. (ed.): The black hole* 118-131.
[hep-th/9309145].
}

\lref\MaldacenaRE{
  J.~M.~Maldacena,
  ``The Large N limit of superconformal field theories and supergravity,''
Int.\ J.\ Theor.\ Phys.\  {\bf 38}, 1113 (1999), [Adv.\ Theor.\ Math.\ Phys.\  {\bf 2}, 231 (1998)].
[hep-th/9711200].
}

\lref\GiribetEQD{
  G.~Giribet,
  ``Scattering of low lying states in the black hole atmosphere,''
Phys.\ Rev.\ D {\bf 94}, no. 2, 026008 (2016), Addendum: [Phys.\ Rev.\ D {\bf 94}, no. 4, 049902 (2016)].
[arXiv:1606.06919 [hep-th]].
}

\lref\fzz{ V.A. Fateev, A.B. Zamolodchikov and Al.B. Zamolodchikov, unpublished.
}

\lref\WakimotoGF{
  M.~Wakimoto,
  ``Fock representations of the affine lie algebra A1(1),''
Commun.\ Math.\ Phys.\  {\bf 104}, 605 (1986).
}

\lref\GiveonCMA{
  A.~Giveon, N.~Itzhaki and D.~Kutasov,
  ``Stringy Horizons,''
JHEP {\bf 1506}, 064 (2015).
[arXiv:1502.03633 [hep-th]].
}

\lref\GiveonDXE{
  A.~Giveon, N.~Itzhaki and D.~Kutasov,
  ``Stringy Horizons II,''
JHEP {\bf 1610}, 157 (2016).
[arXiv:1603.05822 [hep-th]].
}

\lref\Bena{
  R.~Ben-Israel, A.~Giveon, N.~Itzhaki and L.~Liram,
  ``Stringy Horizons and UV/IR Mixing,''
JHEP {\bf 1511}, 164 (2015).
[arXiv:1506.07323 [hep-th]].
}

\lref\ItzhakiRLD{
  N.~Itzhaki and L.~Liram,
  ``A stringy glimpse into the black hole horizon,''
JHEP {\bf 1804}, 018 (2018).
[arXiv:1801.04939 [hep-th]].
}

\lref\GiveonDXE{
  A.~Giveon, N.~Itzhaki and D.~Kutasov,
  ``Stringy Horizons II,''
JHEP {\bf 1610}, 157 (2016).
[arXiv:1603.05822 [hep-th]].
}
\lref\Benb{
  R.~Ben-Israel, A.~Giveon, N.~Itzhaki and L.~Liram,
  ``On the black hole interior in string theory,''
JHEP {\bf 1705}, 094 (2017).
[arXiv:1702.03583 [hep-th]].
}

\lref\GiribetOUF{
  G.~Giribet,
  ``Stringy horizons and generalized FZZ duality in perturbation theory,''
JHEP {\bf 1702}, 069 (2017).
[arXiv:1611.03945 [hep-th]].
}

\lref\MaldacenaHW{
  J.~M.~Maldacena and H.~Ooguri,
  ``Strings in AdS(3) and SL(2,R) WZW model 1.: The Spectrum,''
J.\ Math.\ Phys.\  {\bf 42}, 2929 (2001).
[hep-th/0001053].
}

\lref\DiFrancescoOCM{
  P.~Di Francesco and D.~Kutasov,
  ``Correlation functions in 2-D string theory,''
Phys.\ Lett.\ B {\bf 261}, 385 (1991).
}

\lref\DiFrancescoDAF{
  P.~Di Francesco and D.~Kutasov,
  ``World sheet and space-time physics in two-dimensional (Super)string theory,''
Nucl.\ Phys.\ B {\bf 375}, 119 (1992).
[hep-th/9109005].
}

\lref\MaldacenaKM{
  J.~M.~Maldacena and H.~Ooguri,
  ``Strings in AdS(3) and the SL(2,R) WZW model. Part 3. Correlation functions,''
Phys.\ Rev.\ D {\bf 65}, 106006 (2002).
[hep-th/0111180].
}

\lref\MaldacenaKY{
  J.~M.~Maldacena,
  ``Black holes in string theory,''
[hep-th/9607235].
}

\lref\HawkingSW{
  S.~W.~Hawking,
  ``Particle Creation by Black Holes,''
Commun.\ Math.\ Phys.\  {\bf 43}, 199 (1975), Erratum: [Commun.\ Math.\ Phys.\  {\bf 46}, 206 (1976)].
}

\lref\PeetHN{
  A.~W.~Peet,
  ``TASI lectures on black holes in string theory,''
[hep-th/0008241].
}

\lref\GiveonDXE{
  A.~Giveon, N.~Itzhaki and D.~Kutasov,
  ``Stringy Horizons II,''
JHEP {\bf 1610}, 157 (2016).
[arXiv:1603.05822 [hep-th]].
}

\lref\GiveonICA{
  A.~Giveon and N.~Itzhaki,
  ``String theory at the tip of the cigar,''
JHEP {\bf 1309}, 079 (2013).
[arXiv:1305.4799 [hep-th]].
}
\lref\AttaliGOQ{
  K.~Attali and N.~Itzhaki,
  ``The Averaged Null Energy Condition and the Black Hole Interior in String Theory,''
Nucl.\ Phys.\ B {\bf 943}, 114631 (2019).
[arXiv:1811.12117 [hep-th]].
}

\lref\MaldacenaHI{
  J.~M.~Maldacena,
  ``Long strings in two dimensional string theory and non-singlets in the matrix model,''
JHEP {\bf 0509}, 078 (2005), [Int.\ J.\ Geom.\ Meth.\ Mod.\ Phys.\  {\bf 3}, 1 (2006)].
[hep-th/0503112].
}

\lref\AharonyQU{
  O.~Aharony and E.~Witten,
  ``Anti-de Sitter space and the center of the gauge group,''
JHEP {\bf 9811}, 018 (1998).
[hep-th/9807205].
}

\lref\GiveonUP{
  A.~Giveon and D.~Kutasov,
  ``Notes on AdS(3),''
Nucl.\ Phys.\ B {\bf 621}, 303 (2002).
[hep-th/0106004].
}

\lref\ItzhakiJT{
  N.~Itzhaki,
  ``Is the black hole complementarity principle really necessary?,''
[hep-th/9607028].
}

\lref\BraunsteinMY{
  S.~L.~Braunstein, S.~Pirandola and K.~Zyczkowski,
  ``Better Late than Never: Information Retrieval from Black Holes,''
Phys.\ Rev.\ Lett.\  {\bf 110}, no. 10, 101301 (2013).
[arXiv:0907.1190 [quant-ph]].
}

\lref\MathurHF{
  S.~D.~Mathur,
  ``The Information paradox: A Pedagogical introduction,''
Class.\ Quant.\ Grav.\  {\bf 26}, 224001 (2009).
[arXiv:0909.1038 [hep-th]].
}

\lref\KutasovRR{
  D.~Kutasov,
  ``Accelerating branes and the string/black hole transition,''
[hep-th/0509170].
}
\lref\AharonyAN{
  O.~Aharony and D.~Kutasov,
  ``Holographic Duals of Long Open Strings,''
Phys.\ Rev.\ D {\bf 78}, 026005 (2008).
[arXiv:0803.3547 [hep-th]].
}

\lref\GiribetFY{
  G.~Giribet and C.~A.~Nunez,
  ``Aspects of the free field description of string theory on AdS(3),''
JHEP {\bf 0006}, 033 (2000).
[hep-th/0006070].
}

\lref\fzz{ V.A. Fateev, A.B. Zamolodchikov and Al.B. Zamolodchikov, unpublished.
}

\lref\GiribetKCA{
  G.~Giribet and A.~Ranjbar,
  ``Screening Stringy Horizons,''
Eur.\ Phys.\ J.\ C {\bf 75}, no. 10, 490 (2015).
[arXiv:1504.05044 [hep-th]].
}

\lref\CallanIA{
  C.~G.~Callan, Jr., E.~J.~Martinec, M.~J.~Perry and D.~Friedan,
  ``Strings in Background Fields,''
Nucl.\ Phys.\ B {\bf 262}, 593 (1985).
}

\lref\AtickSI{
  J.~J.~Atick and E.~Witten,
  ``The Hagedorn Transition and the Number of Degrees of Freedom of String Theory,''
Nucl.\ Phys.\ B {\bf 310}, 291 (1988).
}

\lref\ElitzurCB{
  S.~Elitzur, A.~Forge and E.~Rabinovici,
  ``Some global aspects of string compactifications,''
Nucl.\ Phys.\ B {\bf 359}, 581 (1991).
}

\lref\MandalTZ{
  G.~Mandal, A.~M.~Sengupta and S.~R.~Wadia,
  ``Classical solutions of two-dimensional string theory,''
Mod.\ Phys.\ Lett.\ A {\bf 6}, 1685 (1991).
}

\lref\WittenYR{
  E.~Witten,
  ``On string theory and black holes,''
Phys.\ Rev.\ D {\bf 44}, 314 (1991).
}

\lref\DijkgraafBA{
  R.~Dijkgraaf, H.~L.~Verlinde and E.~P.~Verlinde,
  ``String propagation in a black hole geometry,''
Nucl.\ Phys.\ B {\bf 371}, 269 (1992).
}

\lref\MaldacenaRE{
  J.~M.~Maldacena,
  ``The Large N limit of superconformal field theories and supergravity,''
Int.\ J.\ Theor.\ Phys.\  {\bf 38}, 1113 (1999), [Adv.\ Theor.\ Math.\ Phys.\  {\bf 2}, 231 (1998)].
[hep-th/9711200].
}

\lref\KutasovRR{
  D.~Kutasov,
  ``Accelerating branes and the string/black hole transition,''
[hep-th/0509170].
}

\lref\KarczmarekBW{
  J.~L.~Karczmarek, J.~M.~Maldacena and A.~Strominger,
  ``Black hole non-formation in the matrix model,''
JHEP {\bf 0601}, 039 (2006).
[hep-th/0411174].
}

\lref\BenIsraelMDA{
  R.~Ben-Israel, A.~Giveon, N.~Itzhaki and L.~Liram,
  ``Stringy Horizons and UV/IR Mixing,''
JHEP {\bf 1511}, 164 (2015).
[arXiv:1506.07323 [hep-th]].
}

\lref\MaldacenaCG{
  J.~M.~Maldacena and A.~Strominger,
  ``Semiclassical decay of near extremal five-branes,''
JHEP {\bf 9712}, 008 (1997).
[hep-th/9710014].
}

\lref\GiveonMI{
  A.~Giveon, D.~Kutasov, E.~Rabinovici and A.~Sever,
  ``Phases of quantum gravity in AdS(3) and linear dilaton backgrounds,''
Nucl.\ Phys.\ B {\bf 719}, 3 (2005).
[hep-th/0503121].
}

\lref\GiveonGFK{
  A.~Giveon and N.~Itzhaki,
  ``Stringy Black Hole Interiors,''
JHEP {\bf 1911}, 014 (2019).
[arXiv:1908.05000 [hep-th]].
}

\lref\GiveonFU{
  A.~Giveon, M.~Porrati and E.~Rabinovici,
  ``Target space duality in string theory,''
Phys.\ Rept.\  {\bf 244}, 77 (1994).
[hep-th/9401139].
}

\lref\ParikhKG{
  M.~Parikh and P.~Samantray,
  ``Rindler-AdS/CFT,''
JHEP {\bf 1810}, 129 (2018).
[arXiv:1211.7370 [hep-th]].
}

\lref\LindbladNigel{
G.~Lindblad and B.~Nagel,
``Continuous bases for unitary irreducible representations of SU(1,1),''
Annales de l'I.H.P. Physique théorique,  Volume 13 (1970) no. 1,  p. 27-56.
}

\lref\ElitzurCB{
  S.~Elitzur, A.~Forge and E.~Rabinovici,
  ``Some global aspects of string compactifications,''
Nucl.\ Phys.\ B {\bf 359}, 581 (1991).
}

\lref\MandalTZ{
  G.~Mandal, A.~M.~Sengupta and S.~R.~Wadia,
  ``Classical solutions of two-dimensional string theory,''
Mod.\ Phys.\ Lett.\ A {\bf 6}, 1685 (1991).
}

\lref\WittenYR{
  E.~Witten,
  ``On string theory and black holes,''
Phys.\ Rev.\ D {\bf 44}, 314 (1991).
}

\lref\DijkgraafBA{
  R.~Dijkgraaf, H.~L.~Verlinde and E.~P.~Verlinde,
  ``String propagation in a black hole geometry,''
Nucl.\ Phys.\ B {\bf 371}, 269 (1992).
}

\lref\unpublished{A.~Giveon and N.~Itzhaki, unpublished work.}

\lref\tHooftKCU{
  G.~'t Hooft,
  ``On the Quantum Structure of a Black Hole,''
Nucl.\ Phys.\ B {\bf 256}, 727 (1985).
}

\lref\TroostUD{
  J.~Troost,
  ``The non-compact elliptic genus: mock or modular,''
JHEP {\bf 1006}, 104 (2010).
[arXiv:1004.3649 [hep-th]].
}

\lref\TeschnerUG{
  J.~Teschner,
  ``Operator product expansion and factorization in the H+(3) WZNW model,''
Nucl.\ Phys.\ B {\bf 571}, 555 (2000).
[hep-th/9906215].
}

\lref\tHooftFKF{
  G.~'t Hooft,
  ``The black hole interpretation of string theory,''
Nucl.\ Phys.\ B {\bf 335}, 138 (1990).
}

\lref\FukudaJD{
  T.~Fukuda and K.~Hosomichi,
  ``Three point functions in sine-Liouville theory,''
JHEP {\bf 0109}, 003 (2001).
[hep-th/0105217].
}

\lref\GiribetKCA{
  G.~Giribet and A.~Ranjbar,
  ``Screening Stringy Horizons,''
Eur.\ Phys.\ J.\ C {\bf 75}, no. 10, 490 (2015).
[arXiv:1504.05044 [hep-th]].
}

\lref\GiveonJV{
  A.~Giveon and D.~Kutasov,
  ``The Charged black hole/string transition,''
JHEP {\bf 0601}, 120 (2006).
[hep-th/0510211].
}

\lref\HorowitzCD{
  G.~T.~Horowitz and A.~Strominger,
  ``Black strings and P-branes,''
Nucl.\ Phys.\ B {\bf 360}, 197 (1991).
}

\lref\MaldacenaCG{
  J.~M.~Maldacena and A.~Strominger,
  ``Semiclassical decay of near extremal five-branes,''
JHEP {\bf 9712}, 008 (1997).
[hep-th/9710014].
}

\lref\BarsSR{
  I.~Bars and K.~Sfetsos,
  ``Conformally exact metric and dilaton in string theory on curved space-time,''
Phys.\ Rev.\ D {\bf 46}, 4510 (1992).
[hep-th/9206006].
}

\lref\TseytlinMY{
  A.~A.~Tseytlin,
  ``Conformal sigma models corresponding to gauged Wess-Zumino-Witten theories,''
Nucl.\ Phys.\ B {\bf 411}, 509 (1994).
[hep-th/9302083].
}

\lref\AharonyXN{
  O.~Aharony, A.~Giveon and D.~Kutasov,
  ``LSZ in LST,''
Nucl.\ Phys.\ B {\bf 691}, 3 (2004).
[hep-th/0404016].
}

\lref\ElitzurMM{
  S.~Elitzur, O.~Feinerman, A.~Giveon and D.~Tsabar,
  ``String theory on $AdS_3\times S^3\times S^3\times S^1$,''
Phys.\ Lett.\ B {\bf 449}, 180 (1999).
[hep-th/9811245].
}

\lref\GiveonZM{
  A.~Giveon, D.~Kutasov and O.~Pelc,
  ``Holography for noncritical superstrings,''
JHEP {\bf 9910}, 035 (1999).
[hep-th/9907178].
}

\lref\MertensZYA{
  T.~G.~Mertens, H.~Verschelde and V.~I.~Zakharov,
  ``Random Walks in Rindler Spacetime and String Theory at the Tip of the Cigar,''
JHEP {\bf 1403}, 086 (2014).
[arXiv:1307.3491 [hep-th]].
}

\lref\ArgurioTB{
  R.~Argurio, A.~Giveon and A.~Shomer,
  ``Superstrings on AdS(3) and symmetric products,''
JHEP {\bf 0012}, 003 (2000).
[hep-th/0009242].
}

\lref\PolchinskiRR{
  J.~Polchinski,
  ``String theory. Vol. 2: Superstring theory and beyond,''
}

\lref\BarsSV{
  I.~Bars and J.~Schulze,
  ``Folded strings falling into a black hole,''
Phys.\ Rev.\ D {\bf 51}, 1854 (1995).
[hep-th/9405156].
}

\lref\BarsXI{
  I.~Bars,
  ``Folded strings in curved space-time,''
[hep-th/9411078].
}

\lref\BarsQM{
  I.~Bars,
  ``Folded strings,''
Lect.\ Notes Phys.\  {\bf 447}, 26 (1995).
[hep-th/9412044].
}

\Title{
} {\vbox{
\bigskip\centerline{
Stringy Information and Black Holes}}}
\medskip
\centerline{\it Amit Giveon${}^{1}$ and Nissan Itzhaki${}^{2}$ }
\bigskip
\smallskip
\centerline{${}^{1}$Racah Institute of Physics, The Hebrew
University} \centerline{Jerusalem 91904, Israel}
\smallskip
\centerline{${}^{2}$ Physics Department, Tel-Aviv University, Israel} \centerline{Ramat-Aviv, 69978, Israel}
\smallskip

\smallskip

\vglue .3cm

\bigskip

\bigskip
\noindent

We  show
that in string theory, due to non-perturbative effects, there are cases in which two states that semi-classically are completely different,
are in fact the same. One state cannot be excited without exciting the other; they are two components of the same state in the exact theory.
As a result, in some situations that include black holes,
the nature of information in string theory is dramatically different than in field theory.
In particular, each general-relativity state, that lives in the atmosphere of black fivebranes,
is accompanied with an excitation that lives on folded strings, which fill the black-hole interior.
This is likely related  to the way that information is extracted from black holes in string theory,
and we refer to it as stringy information.


\bigskip

\Date{}

\newsec{Introduction and summary}

Very loosely speaking, there are two scenarios, both of which can be traced to the work of 't Hooft from about 30 years ago \refs{\tHooftKCU,\tHooftFKF},
that prevent black holes from destroying information. The brutal and gentle scenarios.
In the brutal scenario, there is a mechanism, yet to be found, that prevents information from falling into the black hole (BH).
In the  gentle scenario, information can fall into the BH, but it is somehow also encoded outside the BH.

Being the leading candidate for a theory of quantum gravity, string theory is expected to indicate which way to go.
So far, we have learned from string theory, via the AdS/CFT correspondence \MaldacenaRE, that BHs cannot destroy information,
but we still do not understand why -- what is the mechanism that encodes the information in the radiation?
Is it brutal or gentle?  The reason why string theory fails to provide an answer so far is that, in general,
we expect non-perturbative effects in the Newton constant, $G_N\sim g_s^2$, where $g_s$ is the string coupling,
to play a key role in resolving the BH information puzzle, and these are hard to describe in string theory.

Black fivebranes are interesting in that regard,
since there is an exact worldsheet conformal field theory (CFT) description of their near horizon regime.
As a result, non-perturbative effects in $\alpha'=l_s^2$,
where $l_s$ is the string length scale,
are well understood.
The exactness of the CFT is useful only for small $g_s$, in which case $l_p \ll l_s$, where $l_p$ is the Planck length scale.
Hence, generally speaking, one would expect the physics associated with the non-perturbative effects in $l_s$ to be quite different than
the non-perturbative effects in $l_p$.
Still, since perturbative effects in $\alpha'$ generate similar terms in the effective action as perturbative effects in $G_N$,
it is natural to wonder if non-perturbative $\alpha'$ effects can teach us something useful about the BH information puzzle.

In this paper, we argue that they do.
We show that, at least in the case of black fivebranes, non-perturbative effects modify the nature of information in string theory
in a rather dramatic fashion;  two states that semi-classically are completely different,
might actually be the same in string theory.
More precisely, they are two components of the same state, which in practice means that they cannot be excited separately, but only simultaneously.
In particular, we show that each of the ordinary general relativity (GR) modes, that propagates in the BH atmosphere,
has a partner in the BH interior. If the GR mode is excited then the partner must be excited too.
This is in the spirit of $A=R_B$ \refs{\SusskindUW,\PapadodimasAQ}  and the ER=EPR \MaldacenaXJA\  proposals,
which suggest a gentle scenario. However, as we shall see, things are more complicated.

What allows us to make such a precise identification is the fact that
the stringy setup that we  inspect is that of near near-extremal black fivebranes,
e.g. those corresponding to a stack of $k$ Neveu-Schwarz (NS) fivebranes.
String theory on this background is described by an exact two-dimensional CFT.
In particular, the physics in the radial and time directions is described by the coset $SL(2,\IR)_k/U(1)$  CFT \refs{\BarsRB\ElitzurCB\MandalTZ\WittenYR-\DijkgraafBA}.
This coset description led to some exact results, that manage to sum up all non-perturbative effects in $l_s$, on the sphere \TeschnerUG,
and in some cases on the torus \TroostUD.

The target-space interpretation of these non-perturbative effects turned out to be quite non-trivial even for large $k$,
when at least naively, $\alpha'$ corrections are expected to be negligible
(see e.g.~\refs{\GiveonCMA\GiveonDXE\GiribetEQD\GiribetOUF-\ItzhakiRLD}, for a review).
For example, the target-space interpretation of the exact reflection coefficient,~\TeschnerUG,
is quite surprising in a rather transparent way,~\refs{\GiveonCMA,\Bena}, in the Euclidean BH case -- the cigar geometry,
and in a more subtle way,~\refs{\Benb,\ItzhakiRLD}, in the Lorentzian case.

The relation between the non-perturbative effects in $\alpha'$ and  stringy information can be traced all the way to the FZZ duality \refs{\fzz,\KazakovPM}. The target-space interpretation of the FZZ duality is the following.  We start with a cylinder background $S^1\times R_{\phi}$, where $R_{\phi}$ is a spatial direction with a linear dilaton, $\Phi(\phi)=-{1\over 2}Q\phi$, and the radius of the $S^1$  is $2/Q$.~\foot{We set $\alpha'=2$ here.} The strong coupling region, at $\phi \to -\infty$,
can be chopped of by condensing an operator, that we refer to as $I$,  that turns the cylinder into the cigar $SL(2,\IR)_k/U(1)$ CFT
background (with $k=2/Q^2$).~\foot{In the supersymmetric case, which we discuss here, it amounts to the continuation of
the black fivebranes above to Euclidean space-time.}
The condensate of $I$ determines the string coupling at the tip, $g_{0}$. The bigger the condensate is the smaller $g_{0}$ is. $I$ is  a truly marginal operator and the target-space interpretation of this is that the free energy associated with this background vanishes. FZZ duality implies that the condensation of $I$ is accompanied
with a condensation of a completely different operator -- the sine-Liouville operator -- which we denote by $W=W^+ + W^-$. The reason for this notation is that from a stringy point of view, the sine-Liouville operator is a linear combination of a string that winds the $S^1$ with a
winding number $\omega=1$ and its conjugate one, with $\omega=-1$.
Regardless of the orientation of the string, its energy is minimized at the tip, and so the wave function of the sine-Liouville operator is heavily suppressed away from the tip.

The target-space interpretation of the FZZ duality is fascinating. At large $\phi$, the wave functions of $I$ and $W^{\pm}$ are very different: $I\sim \exp(-Q\phi)\simeq g_s^2(\phi)$ while $W^{\pm}\sim \exp(-\phi/Q)$. Still, the FZZ duality implies that they amount to the same state or, more precisely, two components of the same state. The two are tied together by the boundary condition at the tip \KarczmarekBW. From the underlying $SL(2,\IR)$ perspective, the origin of the FZZ duality is an isomorphism of representations \GiveonDXE. This implies that the FZZ duality  can be generalized to many other pairs of states. In fact, in \GiveonDXE\ it was shown that all the states that live at the tip of the cigar~\foot{We discuss the cigar with a large curvature length, $\sqrt{\alpha'k}$, a.k.a. a parametrically small $Q$.} -- with a wave function that scales like
$\exp\left[\left(-{1\over Q}+(\ell-1)Q\right)\phi\right]$  ($\ell=1,2,...$) -- have a partner that  lives at the cap of the cigar -- with a wave function that scales like $\exp(-Q\ell\phi)$, respectively.  An observer that has access only to large $\phi$ will  think that these must be two separated states that for some reason get excited simultaneously.
An observer that has access to the tip of the cigar will realize that these are two components of the same state,
which are linked by the boundary condition at the tip.

Upon analytic continuation, the tip of the cigar is mapped to the horizon of the BH and the cap of the cigar to the  atmosphere of the BH, where most Hawking quanta live  before they escape to infinity. Hence, it is natural to wonder if a Lorentzian version of the Generalized FZZ (GFZZ) duality exists and if it gives a precise realization of $A=R_B$ and/or ER=EPR. There is, however, a basic  problem with the analytic continuation of the FZZ duality.  The analytic continuation of $W^{\pm}$ is not mutually local with vertex operators, $V_E$, that are associated with ordinary states that propagate in the BH geometry and carry some finite energy, $E$.

Only very recently, a way out was proposed \GiveonGFK. From the point of view of the underlying $SL(2,\IR)$ theory, the reason why the truly marginal operators $I$ and $W$ condense is that they are invariant under the $SL(2,\IR)_L\times SL(2,\IR)_R$ current algebra; they are screening operators. Simply put, their condensation does not break any of the  symmetries. It is known for many years that there is yet another truly marginal operator, denoted by $F$, that does not break   the $SL(2,\IR)_L\times SL(2,\IR)_R$ current algebra \refs{\BershadskyMF,\GerasimovFI}. $F$ is interesting since, unlike $W^{\pm}$, it is mutually local with $V_E$ and so it can condense also in the BH case. $F$, however, appears to be problematic since it is well defined only for integer $k$, and it is outside the unitarity bound.  This is the reason why to a large extent $F$ was ignored. Nevertheless, in a series of papers \refs{\GiribetFY\GiribetFT\FukudaJD-\GiveonUP}, it was shown that a condensation of the formal analytic continuation of $F$ to any $k$ describes the same non-perturbative physics in $\alpha'$ as the condensation of $W=W^++W^-$.

In \GiveonGFK, it was  proposed that the success of \refs{\GiribetFY,\GiribetFT} is not accidental.
It follows from the fact that $F$ is a fusion of $W^{+}$ with a $W^{-}$, schematically,
\eqn\fffff{F\sim W^+*W^-~,}
namely, $F$ is a bound state of $W^+$ and $W^-$. The calculations done in \refs{\GiribetFY,\GiribetFT}, with $F$ as a screening operator, give the same results as the ``correct calculations," \refs{\FukudaJD,\GiveonUP},
with $W^++W^-$ as a screening operator, since the contribution comes from the points where $W^+$ and $W^-$ coincide.

The target-space interpretation of \fffff\ is in the spirit of the Hartle-Hawking wave function \HartleAI,
but with a stringy twist. The latter involves folded strings, which were proposed in our context in \refs{\ItzhakiGLF,\AttaliGOQ},
and are related to those in \refs{\BarsSV\BarsXI-\BarsQM}.
Concretely,
in the Euclidean section, $W^{+}+W^{-}$ condenses.
In the Lorentzian section, $W^+$ and $W^-$ are glued together to form a
folded string,
$F$~\GiveonGFK; see figure 1.
The upshot is that the eternal BH geometry is accompanied with folded strings that are described by a condensation of $F$.
Just like in the FZZ duality, the eternal BH geometry and the folded strings are two components,
that from the target-space point of view look very different, of the same state.

\ifig\loc{{\it The target-space interpretation of the CFT fusion \fffff:}
In the Euclidean section (on the right), the cigar background is accompanied,
due to the FZZ duality, with a condensate of winding $\omega=\pm 1$ strings, denoted in the text by $W^{\pm}$.
As usual, the Lorentzian section is obtained by cutting the cigar and analytically continue.
When doing so, $W^{\pm}$ are combined to form a folded string, that folds towards the BH.
}
{\epsfxsize4.0in\epsfbox{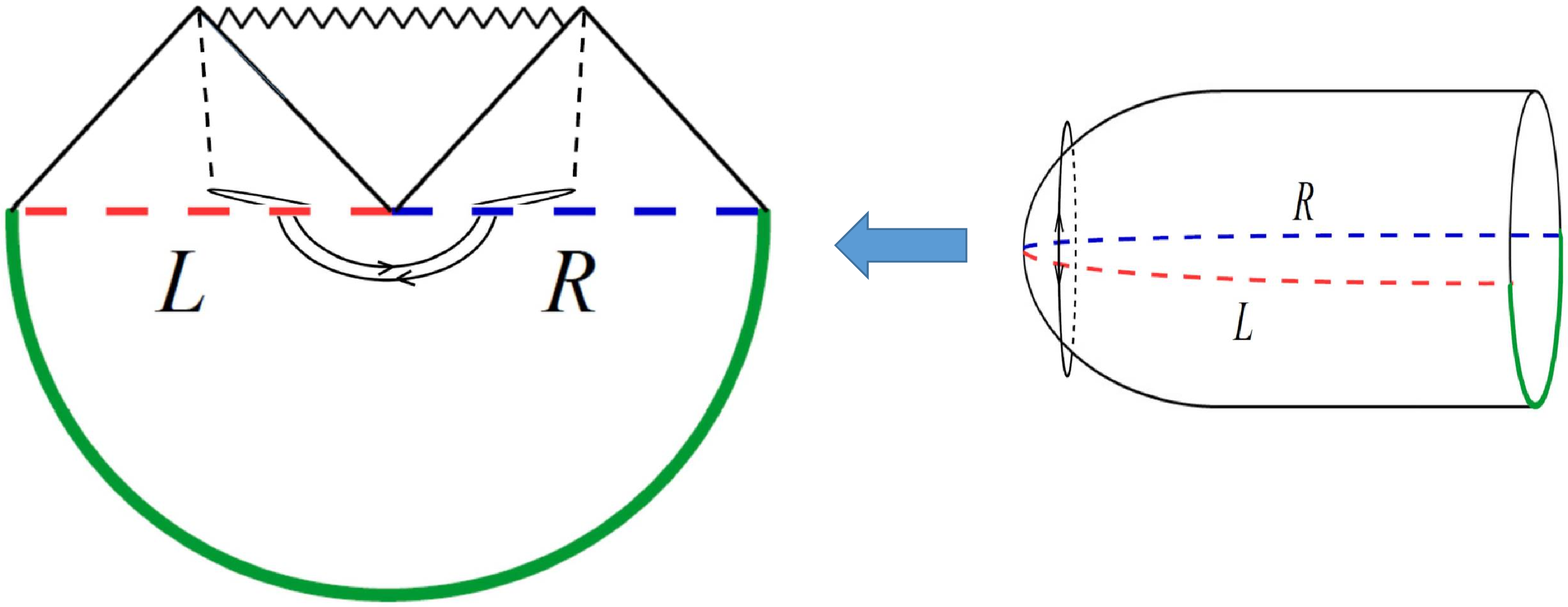}}

The relation with  ER=EPR can be made more precise. Starting with the Euclidean cylinder and following the Hartle-Hawking procedure, we get two disconnected flat space-times. Condensation of $I$ connects them, by forming an ER bridge.
On the other hand, condensation of $F$ does something quite  different.
$F$ is the operator that creates a folded string. The folded string lives on both space-times.
Consequently, a condensation of $F$ entangles the two space-times. A condensation of $F$, therefore, realizes EPR.

This can be extended to include excitations that propagate in the BH atmosphere.
Such states are accompanied by localized dual modes, which live on the folded string,
in the BH interior.
Again, the two are different components of the same state.
One cannot have an excitation in the BH atmosphere without the mode that excites the folded string.
And, again, this follows from isomorphism of $SL(2,\IR)$ representations.
The details  are described in the bulk of the paper; here, we sketch the way this works.

Consider a massless excitation that propagates in the $SL(2,\IR)/U(1)$ BH geometry using the tortoise coordinate, $r^*$. The potential such an excitation experiences vanishes at the horizon, $r^*\to -\infty$, and goes to a constant at infinity,
$r^*\to\infty$. The constant is induced by the linear dilaton and it scales like $Q^2\sim 1/k$. The potential is monotonic with a small mass gap.  From the underlying $SL(2,\IR)$ point of view, the excitations above the gap, $E_s> Q^2$ (where $E_s\sim E^2$ is the energy in the relevant Schrodinger equation
and $E$ is the energy with respect to the Schwarzschild time),
that can escape all the way to $r^*=\infty$, are in the principal continuous representations. As such, they neither have a GFZZ dual,
nor an interior dual.

However, the excitations below the gap, $0<E_s<Q^2$, arise from the principal discrete representations of the underlying $SL(2,\IR)$.
This  is crucial, and at first sight confusing. It is confusing since the energy in the range $0<E_s<Q^2$ is continuous,
still these states are reduced from the discrete representations.
This is crucial, since states in the discrete representations have  GFZZ duals. The GFZZ duals involve spectral flow in the compact,
time-like $J^3$ direction, while $E$ is the energy with respect to the non-compact, space-like $J^2$ direction.
As a result, the details of the GFZZ duals are more subtle,
since they involve a change from the standard basis,  of $J^3$ eigenstates, to the hyperbolic one of $J^2$ eigenstatses. The end result of this procedure is that all the standard states that live in the BH atmosphere are accompanied with a partner that lives on the folded string, in the BH interior, with a
tail outside the BH,
localized a distance of order the string length scale
near the horizon,
and whose wave function scales like~\foot{For parametrically small $Q$, a.k.a for a large BH.}
$\exp\left(-2\phi/Q\right)$, asymptotically in the radial direction.

All of this seems to
fit neatly with  ER=EPR  and/or $A=R_B$. There is, however, a twist in the story.
The original motivation for ER=EPR and $A=R_B$  was to evade the firewall paradox \AlmheiriRT\
(for earlier claims that a unitary BH evolution must result in a singular horizon, see e.g.~\refs{\ItzhakiJT\BraunsteinMY-\MathurHF})
and to extract the information in a gentle fashion.
While the black fivebrane in string theory seems to give a precise realization of the ER=EPR conjecture,
it is far from being clear that it implies a smooth horizon, nor a fuzzy one~\refs{\MathurJK,\AveryTF}.
After all, the fact that $I$ is accompanied with $F$ implies that an infalling observer
will encounter the folded strings and, at least classically,
their energy-momentum tensor  seems to imply a singular horizon~\AttaliGOQ.

Put differently, schematically, the difference between the black fivebranes and ER=EPR is the following.
Indeed, $I$ and $F$ are the operators that correspond to ER and EPR, respectively. However, the  statement is not  that $I=F$. Such a statement would mean that we can either condense $I$ {\it or} condense $F$ and get the same physics.  But this is not the case. For example, the condensation of $I$ is responsible for poles in some correlation functions at imaginary momentum $p\sim iQn$
(with integer $n$), while the condensation of $F$  generates poles in the same correlation functions at $p\sim in/Q$, \GiribetKCA.
Therefore, both  $I$ {\it and} $F$ condense, in an unambiguous way \GiribetFT.
The condensation of $I+F$ means that the ER bridge is not empty, but is filled with folded strings,
which are responsible for the entanglement of the left and right wedges of the eternal BH.

Needless to say that  the fate of an infalling observer  in this setup deserves more serious considerations. In particular, it is possible  that finite $g_s$ effects could render the horizon smooth or fuzzy. This is possible even at arbitrarily small $g_s$, since the number of folded strings is expected to scale like $1/g_s^2$, \AttaliGOQ.

It turns out that this stringy setup
also resonates with the idea that the BH interior is made out of a condensate of gravitons \refs{\DvaliAA,\DvaliRT}.
As discussed above, in the  near-extremal NS fivebranes case, the BH interior is made out of folded strings.
However, the operator $F$, the screening operator that corresponds to the condensation of the folded string,
looks far from the BH like a product of $k$ gravitons: schematically,
\eqn\fti{F=I^k,}
where $I$ is the screening operator that creates the gravitational background.
More precisely, the gravitons bound state dual of the folded string condensate
is the fusion of $k$ gravitons condensates.
We shall refer to such a duality between the BH interior and the bound state of gravitons condensate as BH=GC.

Finally, the black fivebranes also realize the correspondence between BHs and fundamental strings,~\refs{\SusskindWS\HorowitzNW-\GiveonMI}.
The Euclidean version of this was discussed sometime ago \refs{\KutasovRR,\GiveonJV}.
Here, we discuss the Lorentzian case, and emphasis the fact that there is a twist in the plot here too.

\subsec{Outline of the paper}

In the next section, we review some of the properties of black fivebranes.
An important feature of the relevant worldsheet background, near solitonic black fivebranes,
is that it amounts to perturbative string theory on a two-dimensional black hole (BH),
which is described by an {\it exact} worldsheet Conformal Field Theory (CFT) --
the $SL(2,\IR)_k/U(1)$ coset CFT, where $k$ is the fivebranes charge.
Hence, one may reveal exact stringy aspects of the theory,
including those that are {\it non-perturbative} in the string scale, $\alpha'=l_s^2$.

In section 3, we  collect some hints about the nature of stringy information,
from considering the simpler case of string theory on the Euclidean version of the black hole -- the cigar, worldsheet CFT.
We start by reviewing in subsection 3.1 the GFZZ duality.
We define the GR-like modes $I_{\ell,\bar\ell}$, their stringy duals $W_{\ell,\bar\ell}$,
and discuss the reasoning that leads to the conclusion that they are two components of the same state in the theory.
The FZZ duality, which corresponds to $\ell=\bar\ell=1$,
is used in subsection 3.2 to discuss the Euclidean BH-strings transition,
that takes place when the size of the BH is $\ell_s$ (at $k=1$).
In subsection 3.3, the definition of $F$ via the fusion \fffff,
and some of its consequences, are discussed.
In subsection 3.4, we combine the GFZZ duality with the fusion to describe excitations of $F$,
that schematically take the form
\eqn\fff{F_{\ell,\bar{\ell}}\sim W*W_{\ell,\bar{\ell}}^*~.}
The GR modes, $I_{\ell,\bar\ell}$, and their non-perturbative stringy completion, $F_{\ell,\bar{\ell}}$,
can be viewed as the stringy Euclidean version of ER=EPR.
In subsection 3.5, we discuss \fti, and its possible relation with a BH=GC \refs{\DvaliAA,\DvaliRT}.

In section 4, we turn to the actual BH. We start by discussing in subsection 4.1 the subtleties with the analytic continuation
of the winding string condensate, $W$, to Lorentzian signature.
In subsection 4.2, we present the target-space interpretation of the condensate $F$
-- the continuation of the fusion \fffff\ to Lorentzian space-time --
as a BH-filling folded string,
and the light it sheds on the BH-string transition is discussed in subsection 4.3.

We also argue in section 4 that the folded string entangles the two sides of the eternal BH.
This is made more precise in section 5, where excitations of the BH are considered.
It is shown that a GR mode that propagates, say, in the BH atmosphere of the right wedge,
has a non-perturbative stringy completion, in the form of an excitation of the BH interior-filling folded strings,
with tails on both sides of the eternal BH.

Finally, section 6 is devoted for a discussion,
and various technical details are collected in a few appendices.

\newsec{Black fivebranes}

In this section, we describe the ten-dimensional black-brane geometry,
corresponding to Neveu-Schwarz (NS) solitonic branes,
with a finite energy density above extremality,
in  type II superstring theory,
their near horizon limit, that includes  the two-dimensional black hole,
and its powerful description in terms of an exact worldsheet CFT.

The string-frame geometry of $k$ coincident near extremal NS fivebranes in the type II superstring is, \HorowitzCD,
\eqn\aga{ds^2=-\left(1-{r_0^2\over r^2}\right)dt^2+\left(1+{k\alpha'\over r^2}\right)\left({dr^2\over 1-{r_0^2\over r^2}}+r^2d\Omega_3^2\right)+\left(dy_1^2+\dots+dy^2_5\right)~,}
\eqn\agb{e^{2\Phi}=g^2\left(1+{k\alpha'\over r^2}\right)~,}
where $\Phi$ is the dialton field,
and there is an $H$ flux with $k$ units on the three sphere $d\Omega_3^2$,
that we did not write, since it will not play any role below.
This background describes  black fivebranes with mass $M$ per unit five volume $V_5$,
\eqn\agc{{M\over V_5}={M_s^6\over(2\pi)^5}\left({k\over g^2}+\mu\right)~,}
where
\eqn\agd{\mu={r_0^2 M_s^2\over g^2}~;}
$\mu/(2\pi)^5$ is the
energy density above extremality in string units and $M_s$ is the string mass scale,
$M_s=1/l_s=1/\sqrt{\alpha'}~.$

We are interested in the physics near the horizon of the black fivebranes.
The near-horizon geometry is obtained \MaldacenaCG\ by taking the asymptotic string coupling $g$ to zero,
$g\to 0$, while keeping the energy density $\mu$ finite, namely, $r_0\sim gl_s$ in the limit.
Introducing the coordinate $\phi$, related to the radial coordinate $r$ by
\eqn\agf{r=r_0\cosh\left(\phi\over\sqrt{2k}\right)~,}
one finds \MaldacenaCG\ the geometry (from here, we work with $\alpha'=2$)
\eqn\sss{ds^2=-\tanh^2\left(\phi\over\sqrt{2k}\right)dt^2+d\phi^2+2kd\Omega_3^2+\left(dy_1^2+\dots+dy_5^2\right)~,}
and dilaton
\eqn\ddd{e^{2\Phi}={g_0^2\over\cosh^2\left(\phi\over\sqrt{2k}\right)}~,\qquad g_0^2={k\over\mu}~,}
which describes the exterior of a two-dimensional black hole in the time $t$ and radial $\phi$ directions --
a Schwarzschild-like $2d$ black hole \refs{\BarsRB\ElitzurCB\MandalTZ\WittenYR-\DijkgraafBA},
with an asymptotically linear dilaton --
times a three sphere with radius $\sqrt{2k}$ and with $k$ units of $H$ flux,
times a five torus; $g_0$ is the value of the string coupling at the horizon of the black hole,
located at $\phi=0$.

We will also be interested to collect hints from properties of the simpler case of string theory on the Euclidean black hole.
The latter is obtained from the above background by the analytic continuation $t\to ix=i\sqrt{2k}\,\theta$
(and for some purposes, we may Wick rotate also, say, $y_5\to i\tau$, to establish a string theory
on the Euclidean black hole times real time).
The two-dimensional black-hole geometry in \sss\ thus turns into a Euclidean cigar-shaped background, \refs{\BarsRB\ElitzurCB\MandalTZ\WittenYR-\DijkgraafBA},
\eqn\cigar{ds_E^2=2k\tanh^2\left(\phi\over\sqrt{2k}\right)d\theta^2+d\phi^2~,}
with a dilaton given in \ddd.
The angular direction of the cigar, $\theta$, has a standard periodicity,
$\theta\sim\theta+2\pi$,
such that the cigar is smooth at its tip, located at $\phi=0$.
The asymptotic radius of the cigar is $\sqrt{2k}$.
This means, in particular, that the Hawking temperature of the black hole is
\eqn\thaw{T=\beta^{-1}=(2\pi\sqrt{2k})^{-1}~.}
The regime of size of order $\sqrt k$ around the tip is the cap of the cigar;
in this regime, the radius of curvature is also of order $\sqrt k$.
In the Lorentzian case, \sss, it amounts to the thermal atmosphere of the black hole.
On the other hand,
the asymptotic regime is a flat cylinder, $\IR_\phi\times S^1_x$, $x\sim x+2\pi\sqrt{2k}$ ($\IR^{1,1}$ in the Lorentzian case),
with a linear dilaton
\eqn\lindil{\Phi_{asymptotic}=-{Q\over 2}\phi~,}
with a slop
\eqn\qqq{Q=\sqrt{2\over k}~,}
in terms of the black-brane charge, $k$.

The sigma-model worldsheet theories
on both the two-dimensional black-hole background in \sss,\ddd, and in its Euclidean version, \cigar,\ddd,
have very powerful properties that we discuss next.
The geometry \sss,\ddd\ is the sigma-model background of an {\it exact} worldsheet CFT.
Concretely (see e.g. \GiveonFU, for a review),
it is a product of an axial quotient CFT of $SL(2,\IR)$ by its $U(1)$ subgroup in a non-compact space-like direction,
which we denote by the left and right-handed currents $(J^2,\bar J^2)$ of $SL(2,\IR)$,
times the CFT on the group manifold $SU(2)$, times a five-torus, $T^5$,
\eqn\exactcft{{SL(2,\IR)_k\over U(1)}\times SU(2)_k\times T^5~,}
where the level $k$ of both the $SL(2,\IR)$ and the $SU(2)$ affine symmetry algebras is the number of fivebranes above.
The details of the `internal' piece, the $\NN=SU(2)_k\times T^5$ in the particular example \exactcft,
as well as for generic internal spaces in string theory on $SL(2,\IR)_k/U(1)\times\NN$,
will not play a role in the following, and we shall thus ignore it.

Similarly, the Euclidean black-hole background, \cigar,\ddd,
amounts to an exact $SL(2,\IR)_k/U(1)$ quotient CFT, obtained by an axial $U(1)$ gauging of the underlying $SL(2,\IR)_k$ theory
in its compact time-like direction, whose left and right-moving generators we denote by $(J^3,\bar J^3)$, respectively.
In the type II superstring, this worldsheet CFT has an $N=(2,2)$ supersymmetry,
and thus the background \cigar,\ddd\ does not receive perturbative corrections in $\alpha'$ \refs{\BarsSR,\TseytlinMY}.
There are, however, very significant aspects of the theory, which are non-perturbative in $\alpha'$;
they amount to a condensate of a string winding on the angular direction of the cigar, and its excitations.
These are the heroes of the stringy aspects of the Euclidean black hole.
Similarly, their consequences in the Lorentzian case,
are the ingredients that modify dramatically
the GR approximation near the horizon and in the interior of the two-dimensional black hole in string theory.

In the setup discussed above, $k$ is an integer. However, as mentioned, there are
many cases in string theory on $SL(2,\IR)_k/U(1)\times\NN$, in which generically
$k$ is not an integer (see e.g. \refs{\ElitzurMM\GiveonZM-\AharonyXN}).
For example, the near horizon theory of near-extremal I-branes  \ItzhakiTU,
which is built from a stack of $k'$ NS fivebranes stretched in (012345)
intersecting a stack of $k''$  NS fivebranes stretched in (016789), on a circle in $x^1$,
is described by the superstring on the exact ${SL(2,\IR)_k\over U(1)}\times SU(2)_{k'}\times SU(2)_{k''}\times T^2$
(or $\IR^2$) worldsheet CFT,
with ${1\over k}={1\over k'}+{1\over k''}$,~\ElitzurMM.

Finally, as discussed in the introduction, the purpose of these notes is to collect aspects of string theory on the $SL(2,\IR)_k/U(1)$ black hole
and suggest interpretation of their consequences.
The main essence of the physics appears already in the bosonic $SL(2,\IR)/U(1)$ case.
Thus, to avoid presenting the somewhat cumbersome tools involved in the fermionic case,
we will present most of the facts and ideas in the bosonic coset CFT,
though we shall apply their consequences also in the type II superstring context of our ultimate interest, when desired.
In the next section,
we first collect hints from string theory on the Euclidean black hole,
and we shall then turn to the Lorentzian case in section 4.

\newsec{Euclidean hints}

The Euclidean bosonic $SL(2,\IR)_{k_b}/U(1)$ cigar CFT
provides some useful hints about stringy information and its relation with ER=EPR, BH=GC and the BH-string correspondence.
Since, technically, this CFT is much simpler than the eternal BH CFT,
we review in this section some relevant known facts concerning string theory on the cigar,
and add some new ones.

We denote the level of the underlying bosonic $SL(2,\IR)$ theory by $k_b$,
to distinguish it from
\eqn\kkb{k=k_b-2~,}
that appeared previously for the superstring, which is the case we are ultimately interested in.
Many of the details leading to these facts already appear in the literature;
we shall thus refer to existing details, instead of repeating them,
and leave some straightforward calculations, following known manipulations, to the reader.

In the first subsection, we review the GFZZ duality, and
in the second subsection, we discuss the transition at $k=1$, which is related to the BH-string correspondence.
In the third subsection, we discuss the $F\sim W^+* W^-$ relation,
which will play a key role in the Lorentzian case, and in the fourth subsection, we describe its excitations,
which play an important role in a Euclidean realization of ER=EPR.
Finally, in the last subsection, we discuss the relation $F=I^k$, which hints to the BH=GC proposal.

\subsec{Stringy information and the generalized FZZ duality}

In this subsection, we briefly review the GFZZ correspondence studied in \GiveonDXE.
We use the notation of that paper,
apart from $k$ there being the bosonic level $k_b$ here,
and  refer to explicit equations copied from there.

The duality, that is a generalization of the FZZ duality, identifies  states in string theory that   semi-classically are very different.  On one side of the duality we have ordinary states in GR that are bound to the cap of the cigar,
where by ``the cap of the cigar," we mean the region where the curvature is of order $1/k$,
and so its size scales like $\sqrt{k}$ and is much larger than the string size.
These states are denoted by $I_{\ell,\bar\ell}$, with left and right-handed excitation numbers $\ell,\bar\ell\in Z$. On the other side of the duality there are stringy modes (that are simply absent in GR) with winding $\omega=\pm 1$, which we denote by $W_{\ell,\bar\ell}$. These modes are localized near the tip of the cigar.  The precise statement of the duality is that {\it for each $\ell$ and $\bar\ell$ there is only one state in the theory and $I_{\ell,\bar\ell}$ and $W_{\ell,\bar\ell}$ are, semi-classically, two components of the same state.}

Most of what follows relies on this unusual claim, so it is worthwhile to recall its status.
The GFZZ duality was suggested since $I_{\ell,\bar\ell}$ and $W_{\ell,\bar\ell}$ are reduced from operators
that sit in the same representation of the underlying bosonic $SL(2,\IR)$ theory.
A priori, it is possible, of course, that there are more than one state with the same quantum numbers in the theory.
Nevertheless, the possibility that an excitation of $I$ and its $W$ partner amount to different states
is ruled out, e.g. by inspecting the elliptic genus in the supersymmetric extension of the $SL(2,\IR)/U(1)$ theory \GiveonHFA\
(see subsection 7.3 of \GiveonDXE\ for more details) and by the direct, non-trivial study of
correlation functions \refs{\GiribetEQD,\GiribetOUF}.

From the GR point of view, what is special about $I_{\ell,\bar\ell}$ is that these are the only states
that are located at the cap of the cigar at all times. It is related to the fact that they arise from states
in the principal discrete representations of the underlying $SL(2,\IR)$ theory.
There are other states, that come from infinity and spend a finite time at the cap before escaping back to infinity.
These states, that from the $SL(2,\IR)$ point of view sit in different representations
-- the principal continuous ones, do not have GFZZ duals.

For simplicity, in the rest of this subsection,
we present the left-right symmetric operators,
\eqn\iiww{I_\ell\equiv I_{\ell,\ell}~,\qquad W_\ell\equiv W_{\ell,\ell}~,}
and  describe them in some detail.
For completeness, the generic $I_{\ell,\bar\ell}$ and $W_{\ell,\bar\ell}$ are presented in appendix A.

Consider the $SL(2)_{k_b}/U(1)$ cigar CFT. Asymptotically, it is the CFT on the cylinder, $\IR_\phi\times S^1_x$, $x\sim x+2\pi\sqrt{2k}$,
with a linear dilaton \lindil.
Define~\foot{We set $\alpha'=2$ when it is not presented.}
\eqn\www{w={1\over Q}\phi-i\sqrt{k_b\over 2}\tilde x~;\qquad \bar w={1\over Q}\phi+i\sqrt{k_b\over 2}\tilde x~,}
where $\tilde x=x_L-x_R$ is the T-dual coordinate to a canonically normalized scalar field, $x=x_L+x_R$,
and the background charge $Q$ of $\phi$ is related to the `total level' $k$ in \kkb\ as in section 2,  \qqq.

The operators $I_\ell$, that generate the GR bound states in the cigar CFT, are described asymptotically
by the operators in eq. (5.13) of \GiveonDXE,
\eqn\iell{I_\ell\equiv(\beta\bar\beta)^\ell e^{-Q\ell\phi}~,\qquad \ell=1,2,...,\left[{k-1\over 2}\right]~,}
where
\eqn\thus{\beta^\ell=\left(\partial^l e^{-w}\right)e^w=-P_\ell(\partial w,\cdots)~.}
Here, $\beta,\bar\beta$ are induced from the Wakimoto free field description of the underlaying
$SL(2,\IR)$ theory.

The explicit form of $P_\ell(\partial w,\cdots)$ was found in \GiveonDXE\ for $\ell=1,2,3$.
In appendix B, we obtain the general expression  \thus.
From \iell, it is clear that
when the excitation level, $\ell$, is much smaller than the curvature radius, $\sqrt k$,
the wave function of the bound states \iell\ is spread over the whole cap of the cigar.

On the other side of the duality, $W_\ell$ takes the form (see equation~\foot{Some sign conventions are different here.} (5.17) of \GiveonDXE)
\eqn\well{W_\ell\equiv e^{i\sqrt{k_b\over 2}\tilde x-Q\left({k_b\over 2}-\ell\right)\phi}~,
\qquad \ell=1,2,...,\left[{k-1\over 2}\right]~.}
These operators amount to a string winding once around $S^1_x$, with no angular momentum.
The consideration of their anti-winding partners, $W_\ell^*$, is obvious.
It is clear from the exponential factor in $\phi$ that these states are highly localized
around the tip of the cigar,
when $\ell\ll\sqrt k$, namely, when the excitation number
$\ell$ is much smaller than the size of the cap, $\sqrt k$.~\foot{It was shown, \refs{\KutasovRR,\GiveonICA,\MertensZYA,\GiveonHFA},
that they are localized at a distance of order the string length, $l_s$,
around the tip of the cigar, for parametrically small curvature, $Q$.}

A detailed description of the GFZZ duality in the $N=(2,2)$ $SL(2,\IR)/U(1)$ SCFT, which is needed for the black fivebranes, is straightforward,
though cumbersome, \unpublished, but the significant issues for the physics of the Euclidean black hole in string theory
are isomorphic to those in the bosonic coset CFT.

So far, the discussion was at the CFT level.
What does this mean in string theory? For $I_{\ell} \sim W_{\ell}$
to correspond to an on-shell state in string theory, we have to have an extra time direction.
Then, the statement in string theory is that two states, one that behaves asymptotically like $\exp(-Q\ell\phi)$
and another that behaves like $\exp\left[\left(-{1\over Q}+(\ell-1)Q\right)\phi\right]$,
are in fact two components of the same state.
This is, therefore, a situation that does not involve BHs,
in which stringy information is radically different than in quantum field theory on curved space-time.


The case $\ell=\bar\ell=1$ amounts to the FZZ duality between $I$ and $W=W^++W^-$, where
\eqn\iiii{I\equiv I_1\equiv I_{1,1}=\partial w\bar\partial w e^{-Q\phi}~,}
namely, asymptotically on the $\IR_\phi\times S^1_x$ cylinder,
$I$ behaves like the graviton operator~\foot{Up
to a $k$-dependent pre-factor and a total derivative.}
\eqn\iixx{I\sim\partial x\bar\partial xe^{-Q\phi}~,}
and
\eqn\wwww{W^+\equiv W_1\equiv W_{1,1}=e^{-w}~,\qquad W^-\equiv W_1^*=e^{-\bar w}~,}
namely, asymptotically on the cylinder, $W$ is a sine-Liouville operator,
\eqn\sinliu{W\simeq\cos\left({\beta\over 2\pi\alpha'}(x_L-x_R)\right)e^{-{1\over Q}\phi}~,}
where we have reinserted $\alpha'$.~\foot{In the superstring, \sinliu\ is an $N=2$ Liouville superfield,
with $\beta={2\pi\over Q}\sqrt{2\alpha'}$, as in \thaw,
while in the bosonic string $\beta={2\pi\over Q}\sqrt{2(1+Q^2)\alpha'}$.}

\subsec{The $k=1$ transition}


Consider the worldsheet Lagrangian on the cylinder, $ds^2=d\phi^2+dx^2$, with a linear dilaton \lindil,\qqq,
\eqn\lllo{\LL_0=\partial\phi\bar\partial\phi+\partial x\bar\partial x-Q\hat{R}\phi~,}
where $\phi$ and $x$ are canonically normalized free fields.
Both the graviton operator $I$ in \iixx\ and the winding string operator $W$ in \sinliu\
are truly marginal operators in this theory. Hence, adding either of them to $\LL_0$ must give rise to a two-dimensional CFT
which coincides with the linear dilaton one asymptotically.
The interaction $\partial x\bar\partial x e^{-Q\phi}$ is the leading behavior in $1/k$
of a two-dimensional sigma-model on the cigar background, \cigar.
Consequently, adding it to $\LL_0$, the CFT must take care of itself to become the $SL(2,\IR)/U(1)$ cigar theory.
On the other hand, adding $W$ to $\LL_0$ gives rise to a sine-Liouville CFT.
Now, the FZZ duality implies that the latter must take care of itself to become the $SL(2,\IR)/U(1)$ cigar CFT as well.

All in all, the FZZ duality means that the graviton $I$ and the winding tachyon $W$ must condense
simultaneously on the worldsheet, in a correlated way, so that
\eqn\lintiw{\LL_{int}=\lambda_I I+\lambda_W W~,}
with the size of the winding condensate, $\lambda_W$, being related to the size of the graviton condensate, $\lambda_I$,
in an unambiguous way. Indeed, it is known \GiveonUP\ that
\eqn\lwli{\lambda_W={k\over\pi}\left(\pi\lambda_I\Delta(1/k)\right)^{k\over 2}~,}
where
\eqn\deltaz{\Delta(z)\equiv{\Gamma(z)\over\Gamma(1-z)}~.}
The fact that both $\beta$ and $1/Q$ in $W$, \sinliu,
go like $\sqrt k$ at large $k$ means that it is non-perturbative in the $\alpha'$ expansion.
For large $k$, the winding string condensate, $\lambda_WW$ in \lintiw,\sinliu,
decreases rapidly (compared to $I$) as $\phi\to\infty$. This is directly related to the fact that the string winding around the $x$ circle is very heavy there.
Conversely, as one moves to smaller $\phi$, the wave function \sinliu\ grows, and eventually, as one approaches the tip,
one can no longer think of the condensate as describing a wound string. The way to think about it, instead, is discussed in the next subsection.

As $k$ decreases, the falloff of the condensate \sinliu\ becomes less and less rapid.
For $k\to 1$, the wave function of the winding string spreads all the way to infinity.~\foot{The vertex operator
\sinliu\ still falls off at $k=1$, but the corresponding wave function,
which differs from the vertex operators by a factor of $e^\Phi\sim e^{-{Q\over2}\phi}$, does not.}
In this regime, the cigar CFT is better described as a theory on $\IR\times S^1$ with linear dilaton and a sine-Liouville deformation \sinliu.
This transformation of the sigma model on the cigar at large $k$ to a sigma model on the cylinder with a sine-Liouville deformation at small $k$
is an outcome of the FZZ correspondence.
It is an example of a strong-weak coupling duality on the worldsheet.
In general, one should think of the theory as containing both deformations, \lwli\
(the cigar geometry and the thermal winding tachyon); which one is dominant depends on the value of $k$.

As shown in \refs{\GiveonCMA,\GiveonDXE}, and partly reviewed above,
the FZZ correspondence has other consequences, visible already at large $k$.
For example, wave functions of low-energy normalizable states on the cigar,
which typically have an extent $\sqrt{k}l_s$ and are well described by supergravity,
a.k.a. the excitations $I_{\ell,\bar\ell}$ of the graviton condensate,
have another component, the excitations $W_{\ell,\bar\ell}$ of the winding condensate,
which is highly localized near the tip at large $k$.
If one probes the Euclidean BH geometry by scattering,
low-energy probes see the cigar geometry, as expected, while high-energy probes see instead the sine-Liouville one \GiveonDXE.
In particular, the hard wall associated with the fact that the radial direction on a cigar ends at a particular point (the tip),
is replaced for high-energy probes by a soft wall, associated with the sine-Liouville potential \sinliu,
that recedes to more and more negative $\phi$ as the energy of the probe increases.

Thus, non-perturbative $\alpha'$ effects, in particular,
the condensation of the closed string tachyon winding around the Euclidean time direction,
lead to a significant modification of the physics of Euclidean BHs in classical string theory.
These effects are expected to be universal, since Euclidean BHs always contain as part of the geometry a cigar in the radial
and Euclidean time direction, and it is natural to conjecture that a string wrapped around the angular direction of that cigar
has a non-zero condensate in general BH backgrounds.
In particular, we regard the transition at $k=1$, a.k.a. when the size of the BH is the string length scale, $l_s$,
as a (post-)hint to a BH-string transition in string theory: the BH states in string theory on flat
space-time with a linear dilaton cease to exist when $k\leq 1$,
and the density of high-energy states is dominated entirely by perturbative string states.

It was checked \GiveonMI\ that indeed the Bekenstein-Hawking entropy is bigger than that of perturbative string states when $k>1$,
it {\it precisely} matches the one of perturbative string states
in an asymptotically flat space-time with a linear dilaton when $k=1$,
while the latter is bigger than the former when $k<1$.
For other BHs in string theory, e.g. the four-dimensional Schwarzschild BH,
a similar transition is conjectured, \refs{\SusskindWS,\HorowitzNW,\KutasovRR}, when the Schwarzschild radius is of the order of $l_s$.
Although the details of the transition are different -- in the $2d$ Schwarzschild-like case
the order parameter is $k$ while in the $4d$ Schwarzschild case it is the string coupling $g_s$,
the nature of the transition is similar: in both cases the transition occurs when the curvature length
of the cap of the Euclidean BH's cigar is $l_s$.

The main lesson that we take from the above is the following.
Having an exact worldsheet CFT for the two-dimensional BH,
which thus allows us to inspect precisely the non-perturbative effects hinting a BH-string transition,
a.k.a. the winding string condensate,
leads to a (post-)suggestion for similar properties regarding the physics of general BHs in string theory.
Another non-perturbative hint from the Euclidean cigar CFT,
to the nature of the BH-string transition at $k=1$,
will appear in the next subsection.

\subsec{The fusion $F\sim W^{+}*W^{-}$}

In this subsection, we discuss a novel duality that, motivated by the results of \refs{\GiribetFT,\GiveonUP},
was proposed recently in \GiveonGFK.
This duality is key for the understanding of non-perturbative $\alpha'$ effects in the black fivebranes.

The  duality is between the cigar CFT described by the FZZ condensation
\lintiw\ in a linear dilaton theory \lllo,
and the one obtained by
a different condensate,
\eqn\lintif{\LL_{int}=\lambda_I I+\lambda_F F~,}
where $F$ is the fusion of $W^+$ and $W^-$ (defined in \wwww\ with \www). 
The concrete meaning of the fusion is
\eqn\ffff{F(w)\sim\int d^2z W^+(z)W^-(w)~.}
The normalization is such that, for integer $k$,
\eqn\ffffk{F=(\beta\bar\beta)^k e^{-{2\over Q}\phi}~,}
where $\beta^k$ is given in terms of $\phi$ and $x$ in \thus\
and
\eqn\lambdaf{\lambda_F={\left(\pi\lambda_I\Delta(1/k)\right)^k\over\pi\Delta(k)}~,}
where $\Delta(z)$ is given in \deltaz.

Note that $F$ is formally the same as $I_k$ in \iell,
but it
is {\it not}  an excitation of a  graviton.
Since $k>\left[k-1\over 2\right]$, it is outside the unitarity bound and
 it does not generate a state in the theory.
More precisely, it does not generate a single string state in the theory.
Indeed,  \ffff\ implies that it should be viewed as a bound state of $W^{+}$ and $W^{-}$,
which are states in the theory.
Moreover,
$k$ need not be an integer and so generically \ffffk\ is not well defined. We shall get back to this important point.

The precise meaning of this conjectured duality is discussed in \GiveonGFK.
In short, one may calculate correlators
using either  \lintiw\
or
\lintif. As long as the size of the condensate of $F$ is related to that of $W$ by \lambdaf\ combined with \lwli,
one obtains the same result.
Concretely,
inside correlation functions,~\foot{In winding non-conserving correlators,
one uses the winding one, $h=\bar h=0$ degenerate operator, as was proposed in \fzz;
for a review, see e.g. subsection 2.6 in \AharonyXN.}
\eqn\replacef{(\lambda_W)^2 \int d^2z W^+(z)W^-(w)=C_{WF} \lambda_F F(w)~,}
with
\eqn\lwlwclf{(\lambda_W)^2=C_{WF}\lambda_F~,}
that is compatible with \lintif,\lambdaf\ and \lintiw,\lwli,
a.k.a.
\eqn\cwf{C_{WF}=-\pi\Delta(-k)(\lambda_W)^2~,}
which was verified by a direct calculation in \GiveonGFK.
For the two and three-point functions,
this $F-W$ conjecture was checked by combining the results of \GiveonUP\ and \GiribetFT.

Our main physics motivation for such a duality is that it provides a sensible way to think about the
winding condensate of the previous subsection near the tip.
For a large Euclidean BH, the condensate $F$ in \ffff\ (with \thus\ and \www\ inserted) behaves like
\eqn\fffff{F\sim\left(\partial(\phi-ix)\bar\partial(\phi+ix)\right)^k e^{-\sqrt{2k}\phi}+\dots~,}
where the `$\dots$' stand for corrections in $1/k$.
It thus describes a highly excited string that is highly localized near the tip of the cigar.
It is reasonable to think about the winding string condensate in this way:
near the Euclidean horizon, the winding one string condensate, $W^+$, tends to form a bound state with
its conjugate, $W^-$,
giving rise to a highly excited string condensate, instead.

For a small Euclidean BH,
the description of the cigar CFT in terms of \lintif\ adds another ingredient to the BH-string transition
of the previous subsection.
When $k=1$, the graviton operator $I$ and the winding-anti-winding bound-state operator $F$ have the following properties.
First, as can be seen from \iiii\ with \thus\ versus \ffffk,
\eqn\ffii{F=I=\beta\bar\beta e^{-\sqrt{2}\phi}\quad {\rm when}\quad k=1~.}
Second, the operator in \ffii\ is  FZZ dual to a winding operator $W$ whose wave function is marginally non-normalizable
($I$, $W$ and $F$ are at the unitarity bound in the $SL(2,\IR)/U(1)$ CFT when $k=1$; see appendix E).
Moreover, when $k<1$, the winding-anti-winding fused operator $F$ generates a normalizable state.
On the other hand, even though the wave function of the graviton $I$ is normalizable,
the wave function of its FZZ dual $W$ is not normalizable when $k<1$
($I$ and $W$ are outside the unitarity bound in this case; see appendix E).
This means that below the $k=1$ transition point, the description of the worldsheet CFT in terms of the geometry of a cigar
is misleading. Instead, the theory is described in terms of perturbative fundamental strings, which amount to the condensate
$F\sim W*W^*$.

All in all, the FZZ duality
together with the 
fusion, $F\sim W*W^*$, leading to the $F-W$ duality conjecture above,
imply that the tip of the cigar is special already in classical string theory,
due to stringy physics that is non-perturbative in $\alpha'$,
and which can neither be detected via GR considerations, nor taking into account perturbative corrections in $\alpha'$.
We regard it as a Euclidean hint to the possibility that the physics of BH horizons and/or their interiors
is modified dramatically already at the classical level in string theory.
Concretely, it hints to the possibility that the BH interior is a string condensate,
leading to the following.

For small BHs, it hints to a BH-string transition,
as the size of the BH approaches $l_s$,
and when the perturbative string condensate is being `released' from its interior.
For large BHs, we regard the non-perturbative properties of classical string theory on the cigar CFT
as a hint to a realization of a firewall or a fuzzball. This will be discussed in section 4,
but prior to turning to the Lorentzian case, we continue to collect more Euclidean hints.

\subsec{The fusion $F_\ell\sim W*W_\ell^*$}

In subsection 3.1, we reviewed the GFZZ duals of the GR-like excitations $I_{\ell,\bar{\ell}}$
-- the winding one stringy modes $W_{\ell,\bar{\ell}}$ -- in the cigar CFT.
In string theory on the Euclidean black hole (times an extra real time),
while the GR modes are spread in the cigar cap,
their stringy non-perturbative dual partners are localized at the tip.
This suggests \GiveonDXE\ that, at least naively, in string theory
the information in the BH atmosphere is stored also at the BH horizon
(and possibly also in the BH interior, that is absent in the Euclidean geometry).
This exciting possibility will be confirmed in the next sections.

Recall, though, that our motivation for considering string theory on the Euclidean black hole (times an extra real time),
is to collect useful hints regarding the physics of string theory on the Lorentzian BH, in a simple,
well understood setup.
As we shall discuss in section 4,
non-perturbative stringy aspects of perturbative string theory on the Lorentzian BH
have challenging subtleties, yet to be understood.
Those are ameliorated, once we describe the theory in its dual description \lintif.
In the latter framework, the non-perturbative stringy partners of $I_{\ell,\bar{\ell}}$,
which we denote by $F_{\ell,\bar{\ell}}$, are obtained by the fusion of $W_{\ell,\bar\ell}$ with $W$.
So, next, we describe these excitations of $F$, in string theory on the Euclidean BH.

Again, for simplicity we take $\ell=\bar{\ell}$, and consider
the fusion of the winding string condensate $W^+$, \wwww, with the excited winding operators $W_\ell$ in \well,~\foot{The fusion
$F_{\ell,\bar\ell}$ of $W^+$ with generic $W_{\ell,\bar\ell}$ is obvious; it is presented in appendix A.}
\eqn\fusion{F_\ell(w)\sim\int d^2z W^+(z)W_\ell^*(w)~,\qquad \ell=1,2,...,\left[{k-1\over 2}\right]~.}
The normalization is such that, for integer $k$,
\eqn\fffl{F_\ell=(\beta\bar\beta)^{k+1-\ell}e^{-Q(k+1-\ell)\phi}~,}
where $\beta^{k+1-\ell}$ is defined in eq. \thus\ with \www.

The calculation leading to \fusion,\fffl\ is similar to the one done in the particular $\ell=1$ case in \GiveonGFK, mentioned in the previous subsection, and the meaning of \fffl\ for generic $k$ is also similar to that of its low-lying $F=F_1$ condensate, \ffff,\ffffk,
as we discuss next.

The $F_\ell$ in \fusion,\fffl\ are equal formally to the expressions $I_{k+1-\ell}$ in \iell,
but, just like $F$, they are not excitations of the gravity states that are generated by the operators $I_\ell$ in \iell.
The reason is the following.
Recall, \GiveonDXE, that the range of the excitation number $\ell$
for the GR bound states in the cap is $\ell=1,2,...,\left[{k-1\over 2}\right]$.
In the worldsheet CFT on the cigar, the bound $\ell\leq\left[k-1\over 2\right]$ in \iell\
follows from the unitarity bound for such states in affine $SL(2,\IR)_k$.
So, actually, the operators $F_\ell$ in \fusion\ do not generate states in the theory,
since $k+1-\ell>\left[k-1\over 2\right]$,
and thus they are outside the unitarity bound.

Again, just like the ground-state $F$, its excitations $F_{\ell}$
do not correspond to single string states, but to a bound state of $W$ and $W_{\ell}$.
Moreover, generically, $k+1-\ell$ is not an integer, in which case \fffl\ is
a notation for its definition in eq. \fusion.
Indeed, for generic $k$,  the $F_\ell$ in \fusion\ can be thought of as a smeared average over off-shell string excitations,
which collapses to an on-shell excitation for integer $k$.
Note that this is in harmony with the analytic structure of the condensate size, $\lambda_F$, in eqs. \lintif--\replacef:
one may regard the function $\lambda_F(k)$, \lambdaf, as a measure of the smearing, in particular, it develops a pole when $k$ is integer,
in harmony with the collapse of the condensate $F$ and its excitations $F_\ell$ to  on-shell operators, \fffl, in that case.

To recapitulate, the operators $F_\ell$ in eqs. \fusion,\fffl\ are
the non-perturbative partners of $I_{\ell}$.
They describe excited modes of the fused condensate, $F\sim W*W^*$, and are highly localized near the tip of the cigar (as long as the excitation number, $\ell$, is sufficiently smaller than the size of the cap, $\sqrt k$).

The scaling dimension of $F_\ell$ is identical, by construction, \fusion,
to that of $W_\ell$ in \well\ and thus also to $I_\ell$ in \iell,
\eqn\hhhhh{\Delta(\ell)=\ell-{\ell(\ell-1)\over k}~,}
as it must.
Note  that $\Delta(\ell)\geq 1$, and $\Delta=1$ only for $\ell=1$.
Therefore, when considering string theory on the cigar,
$I_\ell$, $W_\ell$ and $F_\ell$, with $\ell>1$, can appear as on-shell physical states only if we have an extra time direction.

 Nevertheless,
we regard the duality between the excitations of the fused condensate near the tip
and the gravity bound states in the cap as a Euclidean hint to a duality between the modes in the interior of the BH
and those in its thermal atmosphere.
This will be clarified in sections 4 and 5, but prior to turning to the Lorentzian case,
we present another intriguing Euclidean hint in the next subsection.

\subsec{ $F=I^k$ and BH=GC}

In this subsection, we shall argue that, in a certain sense,  $F$ may be regarded also as a bound state of $k$ gravitons,
localized near the tip of the cigar. Concretely, for integer $k$, the operator $F$ in \replacef\ takes the form \ffffk,
which can be written schematically as
\eqn\ffik{F=\left(\beta\bar\beta e^{-Q\phi}\right)^k=I^k~;}
this follows from \ffffk\ versus \iiii\ with \thus.

More precisely,
it was argued in \GiribetFT\ that, inside correlators,
\eqn\liklf{(\lambda_I)^k\left[\prod_{i=1}^{k-1}\int d^2 z_i I(z_i)\right]I(w)=C_{IF}\lambda_F F(w)~,}
with
\eqn\likcif{(\lambda_I)^k=C_{IF}\lambda_F~,}
for the $\lambda_F$ in \lintif--\lambdaf.

There is a problem with eq. \liklf: the entire discussion is valid for integer $k$, but for such $k$'s  $\lambda_F(k)$
diverges.\foot{A possible
interpretation of this divergence is discussed in the previous subsection.}
Still, with manipulations that involve, in particular, continuation in $k$,
in \GiribetFT, it was shown that the two and three point functions calculated with the screening operator $F$
are identical to those calculated with the screening operator $I$,
if \liklf\ is satisfied with \likcif,\lambdaf.

Despite the fact that we do not have a physical understanding of these manipulations and the fact that,
unlike $F\sim W *W^{*}$, \liklf\ is defined only for integer $k$,
we find it intriguing, especially since this seems to resonate with ideas presented in \refs{\DvaliAA,\DvaliRT}.
There, it was argued that BHs are filled with a condensate of gravitons.
In the next section, we argue that $F$ is the operator that corresponds to a BH-filling folded string.
Consequently, the relation in eq. \liklf\ implies that,
at least far from the BH, the folded string might be viewed as a bound state of $k$ gravitons.

\newsec{Lorentzian physics}

In this and the next sections,
we consider the Lorentzian black fivebranes,
and see how the Euclidean hints from the previous section turn into rather concrete statements,
regarding what we refer to as stringy information.

In the first subsection, we present the subtleties associated with the continuation of $W$ to the Lorentzian case.
In the second subsection, we turn to the physics of $F$, which avoids these subtleties,
and argue that it amounts to an interior-filling folded string.
Finally, in the last subsection, we discuss the light it sheds on the BH-string transition.

In particular, we argue in this section that $F$ entangles the two sides of the eternal BH,
similar to ER=EPR, whose precise realization as stringy information is clarified
when considering the excitations of the stringy BH, in the next section.

\subsec{$I$ and $W$}

As discussed in section 2,
the Lorentzian $SL(2,\IR)_k/U(1)$ geometry can be obtained e.g. by the analytic continuation
\eqn\anacon{ix\to t~.}
In particular, after analytic continuation, the graviton operator \iixx\ becomes
\eqn\ilorentz{I\sim\partial t\bar\partial t e^{-Q\phi}~,}
up to a total derivative.
This is in harmony with the behavior of the two-dimensional BH geometry in \sss, at asymptotically large radial $\phi$.
And, as in the Euclidean case, once $I$ is condensed in the $\IR_\phi\times\IR_t$
theory with a linear dilaton --
the continuation of \lllo, the theory `takes care' of itself and becomes the Lorentzian $SL(2,\IR)_k/U(1)$ quotient CFT.

The coordinates $-\infty\leq t\leq\infty$ and $\phi\geq 0$ cover the right wedge of an eternal BH;
the former can be extended to the latter geometry in the standard way, e.g. by using Kruskal-Szekeres coordinates,
$-\infty\leq u,v\leq\infty$
(see e.g. \GiveonFU, for a review),
\eqn\vvuu{v=\sinh\left({2\pi\phi\over\beta}\right)e^{2\pi t/\beta}~,\qquad u=-\sinh\left({2\pi\phi\over\beta}\right)e^{-2\pi t/\beta}~,}
in terms of which the two-dimensional BH metric and dilaton in \sss\ and \ddd\ take the form
\eqn\dsvu{ds^2=-{dudv\over 1-uv}~,\qquad e^{2\Phi}={g_0^2\over 1-uv}~,}
respectively.

The $\beta$ in \vvuu\ is the inverse Hawking temperature of the BH.
Recall that in the bosonic $SL(2,\IR)_{k_b}/U(1)$ and fermionic $SL(2,\IR)_k/U(1)$ theories,
$\beta$ is related to the bosonic level, $k_b=k+2$, and the total one, $k$, by
\eqn\bbeta{\beta=2\pi\sqrt{\alpha'k_{b}}\quad ({\rm bosonic})~;\qquad  \beta=2\pi\sqrt{\alpha'k}\quad ({\rm fermionic})~,}
respectively (recall also that we set $\alpha'=2$, when it is not presented explicitly).

The analytic continuation \anacon\ of the sine-Liouville operator \sinliu\ gives~\foot{As in section 3, we present the bosonic case;
similar manipulations can be done straightforwardly in the fermionic case of our ultimate interest, giving rise to the same result.}
\eqn\sll{W\to\,\cosh\left(\sqrt{k_b\over 2}\left(t_L-t_R\right)\right)e^{-{1\over Q}\phi}~.}
Since $W$ arises from an $SL(2,\IR)_L\times SL(2,\IR)_R$ invariant operator in the underlying $SL(2,\IR)$ theory,
it must survive any gauging,
in particular, the one leading to the Lorentzian $SL(2,\IR)/U(1)$ CFT.
This, however,  seems to lead to some non-conventional, apparently inconsistent properties.

For instance, consider the following
operator product expansion (OPE):
\eqn\egope{e^{iEt(z,\bar z)}e^{\sqrt{k_b/2}(t_L-t_R)}\sim
e^{\vartheta\sqrt{2k_b}E}e^{i(E-i\sqrt{k_b/2})t_L+i(E+i\sqrt{k_b/2})t_R},}
where here $t(z,\bar z)=t_L(z)+t_R(\bar z)$, $t_L=t_L(0),\, t_R=t_R(0)$, and $z=|z|e^{i\vartheta}$.
The expression \egope\ appears within OPEs of operators corresponding to generic energy states $V_E$
in the BH CFT with the screening operator $W$ in \sll.
This leads to non-locality on the worldsheet, since upon rotating $z$ around the origin, $\vartheta\to\vartheta+2\pi$,
the r.h.s. of \egope\ picks up a factor of $e^{\beta E}$, with the $\beta$ in eq. \bbeta.

We should note, however, that we are considering an embedding of a Euclidean worldsheet in a non-trivial Lorentzian space-time.
So, it is possible that the non-locality puzzle above is resolved when one considers Lorentzian worldsheets in the eternal BH target.
However, at the moment, it is not known to us how to do it directly.
Nevertheless, for Euclidean worldsheets in the Lorentzian BH background,
we suggest that this should be resolved in the following way.
Going around an insertion of \sll\ on the worldsheet, amounts in space-time to taking
\eqn\ttt{t \to t+ i\beta~.}
The eternal BH coordinates,
$u$ and $v$ in \vvuu, are invariant under \ttt,
and since \egope\ is invariant when combining \ttt\ with
\eqn\nowunder{\vartheta\to\vartheta+2\pi~,}
it is natural to suspect that the apparent non-locality on the worldsheet is not a `bug,'
but rather a `feature' of the eternal BH theory,
which reflects the fact that the BH has a temperature.

In appendix C, we show how \ttt\ and \nowunder\ come about when considering the Lorentzian $AdS_3$ with a Euclidean worldsheet.
Since the Lorentzian $AdS_3$ CFT is
expected to be a consistent target space for perturbative strings,
this is a strong support that the issue is with attempting to describe
the physics of strings in a non-trivial Lorentzian background using a Euclidean worldsheet.


Luckily, as reviewed in the previous section, in order to describe the non-perturbative $\alpha'$ corrections in the BH case,
we do not have to deal with these issues. Instead, we can use the dual description in which $W^{\pm}$ do not condense, but $F$ does.
Since $F$ is mutually local with $V_E$, this dual description appears to be much simpler.
And so next, we turn to discuss the physics of $F$ in the eternal BH theory.

\subsec{
$F$, folded string and  EPR}

The analytic continuation of $F$ is the fusion, \ffff, of the analytic continuation of $W^{\pm}$,
\eqn\wwwwpm{W^\pm=\exp\left(\pm{\beta\over 2\pi\alpha'}(t_L-t_R)\right)e^{-{1\over Q}\phi}~,}
where $\beta$ is given in \bbeta.
As is clear from \ffff\ and \wwwwpm, $F$ is mutually local with respect to $V_E$,
and thus there are no subtleties associated with its condensation.
This implies that  $F$  should have a clear interpretation in the BH geometry.

The goal in this subsection is  to discuss the physical meaning of $F$.
We claim that the semi-classical meaning of $F$ is of a folded string that is filling the entire BH.

The first evidence for this comes from the target-space interpretation of $F\sim W^{+} * W^{-}$, as presented in figure 1.
There, we see how the Hartle-Hawking wave-function procedure fits neatly with $F\sim W^{+} * W^{-}$,
provided that $F$ is associated with a folded string that folds towards the BH, as discussed in \ItzhakiGLF.
Classically, the folded string can fold towards the BH only behind the horizon \refs{\MaldacenaHI,\ItzhakiGLF}.
Outside the BH, the probability for the folded string to fold towards the BH is exponentially suppressed,
as reflected by the $\phi$ dependence of $W^{\pm}$ and $F$,~\GiveonGFK.
The fact that $W^{\pm }$ and $F$ are related via analytic continuation can also be seen at the classical level,
as  illustrated in appendix C.

The argument presented above  is valid for any $k$.
In the special cases of integer $k$, another, more direct, argument can be made.
The asymptotic behavior of the operator $F$ in the Lorentzian BH is obtained by the analytic continuation, \anacon,
of eqs. \ffffk,\fffff, giving rise to
\eqn\florentz{F\sim\left(\partial(\phi-t)\bar\partial(\phi+t)\right)^ke^{-{2\over Q}\phi}+\dots~,}
where the `$\dots$' stand for corrections in $1/k$, which can be read from the analytic continuation \anacon\ of \ffffk\ with \thus\ and \www.
Equation \florentz\ implies that far from the BH, $F$ describes a level $k$ on-shell excitation of a string whose  size is $\sqrt k$.
Now, since  the size of the BH is also $\sqrt k$ and since  \florentz\ implies that the   wave function of the string is highly suppressed away from the BH, it is natural to interpret $F$ as  a folded string that semi-classically is filling the BH interior. This is also reflected in $\lambda_F(k)$, which develops a pole for integer $k$. For non-integer $k$, \florentz\ does not hold, and $\lambda_F(k)$ is regular.
Since \fusion\ is still valid, we expect the size of the string to be $\sqrt k$ in average.

A third evidence comes from the fact that $F$ is a screening operator, a.k.a. it is reduced from a truly marginal operator with
$J^a=\bar J^a=0$ in the underlying $SL(2,\IR)$ theory.
This agrees with the observation made in \refs{\ItzhakiGLF,\AttaliGOQ}
that the BH-filling folded string does not break any of the BH background symmetries.

In light of these arguments, we assume in the rest of the paper that $F$ is the operator that corresponds to the folded strings discussed in \refs{\ItzhakiGLF,\AttaliGOQ},  and see what   physics follows from this assumption.

One  conclusion is that the black fivebranes system realizes the ER=EPR proposal, \MaldacenaXJA, albeit with a twist.
To see this, we recall that $I$ is the operator that creates the ER bridge between the two asymptotic regions.
As we just argued, $F$ corresponds to a folded string. This folded string is an extended object with tails on both asymptotic regions,
and so it entangles the right and the left sides of the BH. It is natural, therefore, to relate $F$ with the EPR side of the duality.
The fact that the folded string entangles the left wedge with the right wedge is made more precise in the next section.
There, we consider excitations of the eternal BH. We show that an excitation that lives on, say, the right wedge of the BH,
is accompanied with an excitation of the folded string with tails both on the right and the left wedges.

The twist is that the CFT statement is {\it not} that $I=F$ (which is the analog of ER=EPR).
Operationally, such a statement would mean that one can {\it either} work with the screening operator $I$ {\it or} with the screening operator $F$.
As discussed above, the correct statements is that both $I$ {\it and} $F$ must condense.\foot{For example,
$I$ and $F$ are responsible for different poles in correlation functions \GiribetKCA.}
Namely, the condensate is of $\lambda_I I+\lambda_F F$. In other words, the ER bridge is not empty, but is filled with folded strings.

Before considering, in the next section,
excitations of the BH and folded string,
which strengthen the above picture about stringy information,
we first comment about the BH-string transition \refs{\SusskindWS,\HorowitzNW} for black fivebranes \refs{\GiveonMI,\KutasovRR},
in the next subsection.

\subsec{$k=1$ and the BH-string transition}

The semiclassical description of the condensate $F$ in terms of an interior-filling folded string
sheds new light on the nature of the transition at $k=1$, discussed in the Euclidean case in subsections 3.2 and 3.3.
In particular, it clarifies its interpretation as realizing the BH-string transition, for the $SL(2,\IR)_k/U(1)$
Lorentzian BH. Next, we discuss it in some detail.

For parametrically large $k$, the folded string is trapped in the interior of the BH, with a tail, $W$, highly localized
at a distance $l_s$ from the horizon. The geometric description of the BH interior is thus misleading;
an infalling observer will encounter the strings condensate as s.he approaches a distance $l_s$ from the horizon.
Even as $k$ decreases, as long as $k$ is large, the change in the effect of the strings condensate outside the BH is hardly felt.
However, as $k\to 1$, the effect of the folded string changes dramatically.
The range of its tail is expanded in the radial direction all the way to $\phi\to\infty$.

For $k<1$, two things happen at once. On the one hand, as in the Euclidean case, the graviton condensate, $I$,
which behaves asymptotically like $e^{-Q\phi}$, has $j+1=1$, and is thus
outside the unitarity bound: $j=0>{k-1\over 2}$ (see appendix E).
Consequently, the description of the the $SL(2,\IR)_k/U(1)$
theory in terms of the geometry of a BH, as in eq. \sss\ and/or eq. \dsvu, is misleading
-- the BH is not a state in the $\IR_t\times\IR_\phi$ theory with a linear dilaton.
On the other hand, the fundamental strings condensate, $F$, which behaves asymptotically like $e^{-Qk\phi}$,
has $j+1=k$ (see appendix E), and is thus inside the unitarity bound when $1/2<k<1$: $-{1\over 2}<j=k-1<{k-1\over 2}$.
Note that the lower bound on $k$ follows from the fact that in the superstring,
the central charge of the $SL(2,\IR)_k/U(1)$ worldsheet SCFT is $c=3+{6\over k}$,
and thus it is over critical, $c>15$, when $k<1/2$.

The transition point, $k=1$, is very special.
At this point, both the fundamental strings screening operator and the gravitons screening operator
are at the boundary of the unitarity bound;
they both have $j=0={k-1\over 2}$. And, actually,
as in the Euclidean case, \ffii, at this point, the operators $F$ in \florentz\ and $I$ in \ilorentz\ coincide,
\eqn\ilfl{F=I\sim\partial t\bar\partial t e^{-\sqrt{2}\phi}~,}
up to a total derivative.

The picture that emerges from the above is thus the following.
At the transition point, $k=1$, the localized folded strings, which, in a sense, replace the BH interior,
are being `released,' as perturbative strings. To recapitulate, while for $k>1$, the better description of the $SL(2,\IR)_k/U(1)$
Lorentzian theory is in terms of a BH background, whose interior is being replaced by folded strings,
for $k<1$, the BH disappears, but the strings remain.
This provides a nice realization of the BH-string transition, in the two-dimensional case.

\newsec{ BH and stringy information }

In this section, we discuss excitations of the eternal BH.
We show that excitations of ordinary GR modes that are confined to the BH atmosphere are accompanied with
excited folded string modes that live mostly inside the BH.
We discuss the implications of this observation.

Let us  start by reviewing some well known facts \DijkgraafBA\ about ordinary GR-like modes that propagate in the
$SL(2,\IR)_k/U(1)$ BH atmosphere. Consider a massless minimally coupled scalar field, $a$, that propagates in the eternal BH geometry that
takes the form, \sss,\ddd,
\eqn\gebh{ds^2=-\tanh^2(\phi/\sqrt{2k})dt^2+d\phi^2,~~~~e^{2\Phi}=g_0^2\cosh^{-2}(\phi/\sqrt{2k}).}
Defining $\Psi(\phi)=a(\rho) \cosh(\phi/\sqrt{2k}) e^{-i E t}$ and performing the coordinate transformation
$\rho=\sqrt{2k}\log(\sinh(\phi \sqrt{2k}))$, we obtain a Schrodinger-like equation in the tortoise coordinates, $\rho$,
\eqn\sch{\left(-\partial^2_\rho + V(\rho)\right) \Psi(\rho)=E^2 \Psi(\rho),}
with
\eqn\potee{V(\rho)={1 \over 2 k} \left(1-{1 \over (1+\exp(2\rho/\sqrt{k} ) )^2}\right) .}
Note that, much like  in Schwarzschild BH, the size and hight of $V(\rho)$ are fixed by the curvature, $1/k$ in our case.
However, here, unlike in Schwarzschild BH, due to the linear dilaton, $V(\rho)$ is a monotonic function.~\foot{This is a
consequence of considering the near fivebrane theory, \sss;
in the full background of the black fivebrane, \aga, $V(r)$ is a step function, as in Schwarzschild BH.}

This implies that there are two kinds of modes (see figure 2).
Modes with $E^2>1/2k$ oscillate both at infinity, $\rho\to \infty$, and at the horizon,
$\rho\to -\infty$. We refer to such modes as scattering modes.
There are also bound state modes, with $1/2k> E^2>0$; these modes are oscillating at the horizon and decay exponentially fast at infinity, and so they are indeed bounded to the BH.~\foot{In Schwarzschild BH, the situation is similar;
the only difference is that the modes in the atmosphere can tunnel to infinity, as the potential goes to zero there.}
From the underlying
$SL(2,\IR)$ point of view, the scattering states arise from the principal continuous representations,
while the bound states arise from the principal discrete representations.
To see this, we recall how these, and other modes that propagate in the $SL(2,\IR)/U(1)$ BH geometry, are described in string theory.

Consider, for example,  type II string theory on $SL(2,\IR)_k/U(1)\times\NN$, namely, on the two-dimensional BH
times an `internal space' $N=2$ SCFT, $\NN$.
For instance, in the theory of $k$ near-extremal NS fivebranes on a five torus, discussed in section 2,
$\NN=SU(2)_k\times T^5$ (or $\IR^5$, in the special case of a non-compact black fivebrane).
A class of physical vertex operators of interest, in the $(-1,-1)$ picture, are (see e.g. \PolchinskiRR, for a review and definitions)
\eqn\vje{V=e^{-\varphi-\bar\varphi}V_{jE}V_\NN~,}
where $V_{jE}$ is a primary operator in the $SL(2,\IR)_k/U(1)$ SCFT,
whose scaling dimension is
\eqn\vjejjee{h(V_{jE})=-{j(j+1)\over k}-{E^2\over 2}~,}
with the energy, $E$, related to the eigenvalue $m_2$ of $J^2$ in the underlying $SL(2,\IR)_k$ theory by
\eqn\eqm{E=Qm_2~,}
and similarly for $\bar h$ (see e.g. \refs{\DijkgraafBA,\GiveonFU}, for a review). The on-shell condition reads:
\eqn\onshellej{-{j(j+1)\over k}-{E^2\over 2}+N-{1\over 2}=0~,}
where $N$ is the left-handed dimension of $V_\NN$, which corresponds to a transverse left-handed string excitation in $\NN$.
A similar condition is obtained for the right-moving excitations, $\bar N(=N)$.

The modes that satisfy \sch\ correspond to $N=\bar{N}=1/2$. There are two ways to satisfy \onshellej\ with $N=1/2$.
Modes with $E^2>1/2k$ have
\eqn\jjjsp{j=-{1\over 2}+is~,}
where $s$ is related to the momentum in the radial direction, $P$, by
\eqn\pqs{P=Qs.}
These are the scattering modes, which according to \jjjsp\ are reduced from the principal continuous representations (see figure 2).

\ifig\loc{{\it The potential experienced by GR-like modes in tortoise coordinates:}
The potential is monotonic with a mass gap that scales like $1/k$.
There are modes, marked in purple, that can escape to infinity,
and modes, marked in red, that cannot.
From the underlying $SL(2,\IR)$ point of view they are different:
the red modes arise from the principal discrete representations and are accompanied by a non-perturbative partner in the BH interior.
The purple modes arise from the principal continuous representations and are not accompanied by a non-perturbative partner.}
{\epsfxsize3.0in\epsfbox{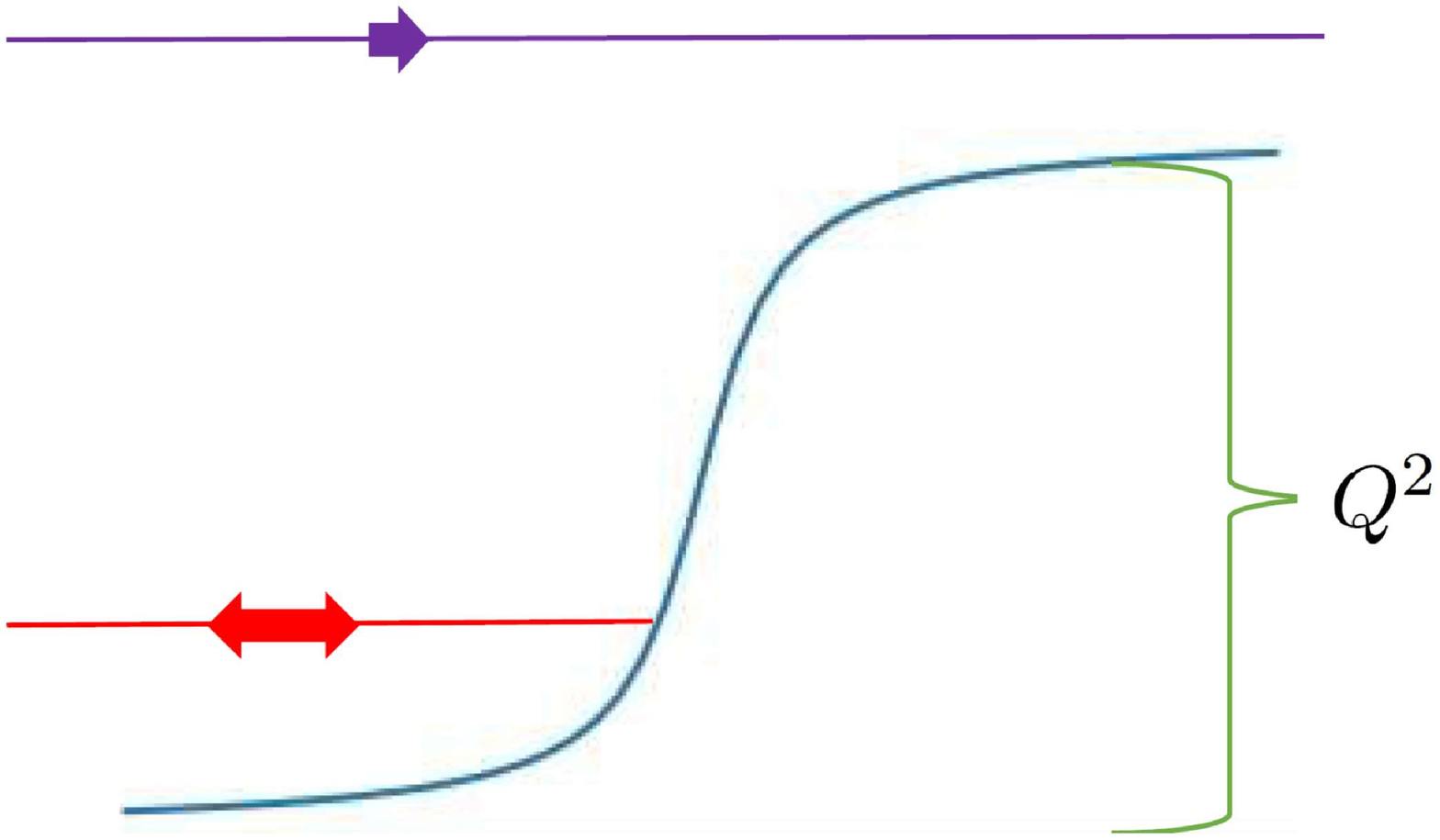}}

There are also modes below the mass gap, with $E^2<1/2k$. Equation \onshellej\ with $N=1/2$ implies that such modes have real $j$.
Namely, they are reduced from the principal discrete representations
(for a review on $SL(2,\IR)$ representations, see e.g. \MaldacenaHW).
States $V$, \vje, that are reduced from the principal discrete representations of $SL(2,\IR)$
are accompanied with a GFZZ dual, that we denote by $W_V$.


To construct $W_V$, we first recall that the operators $V_{jE}$ are reduced via coset decomposition from $J^2$ (and $\bar{J^2}$) eigenstates
in the underlying theory, hence, we work in the hyperbolic basis of $SL(2,\IR)$, and denote such states by $V_{j;m_2}$,
where $m_2$ is the value of $J^2$ (and $\bar m_2=m_2$ below), which is related to the energy $E$ by eq. \eqm.
Each $V_{j;m_2}$ can be written as an infinite sum over states in the standard basis, $V_{j;m_3}$,
where $m_3$ is the value of $J^3$.

Now, each $V_{j;m_3}$ has a known GFZZ dual and, consequently, so is each $V_{j;m_2}$,
whose dual we shall denote by $\Phi_{\tilde j;m_2}^{\omega=1}$. Reducing the latter to the $SL(2,\IR)/U(1)$ coset, will give rise to
the GFZZ duals of GR bound states $V_{jE}$, which we denote by $V_{\tilde jE}^{\omega=1}$.
Together with the contribution of the internal space, we shall thus obtain a physical dual, $W_V$,
to each on-shell GR bound state $V$ in \vje,
\eqn\vvwwje{W_V=e^{-\varphi-\bar\varphi}V_{\tilde jE}^{\omega=1}V_\NN~.}
More details of this construction are presented in appendix D.

Even without the detailed construction of $W_V$,
it is clear that, much like $W$, \wwwwpm, far from the BH, the vertex operators $W_V$, \vvwwje, include  an
$\exp\left(\sqrt{k \over 2}\left(t_L-t_R\right)\right)$ factor.
Hence, they are not mutually local with standard vertex operators, e.g. $V$ in \vje,
that include an $\exp(-i E(t_L+t_R))$ factor, and which are clearly in the theory.
To overcome this issue, we construct $F_V$, that is the Lorentzian analog of the fusion $F_\ell\sim W*W_\ell^*$ 
in the Euclidean case, \fusion; explicitly,
\eqn\fusionv{F_V(w)\sim\int d^2zW^+(z)W_V^*(w)~,}
where the operator $W^+$ is the one in eq. \wwwwpm, and the operators $W_V$ are the GFZZ duals of $V$ discussed above, \vvwwje.

The operators $F_V$ are the non-perturbative completion of $V$.
Each $V$ and its corresponding $F_V$ are two components of the same state in the theory.
$V$ has a clear GR-like interpretation and $F_V$ does not;
it is a stringy mode -- an excitation of the folded string -- that lives mainly behind the BH horizon.
Just like $F$, the operators $F_V$ have a tail that falls rapidly outside the BH.

There is something intriguing about this tail that reveals much of the physics associated with the folded string. Suppose that $V$ is an excitation that appears on the right wedge of the BH and not on the left wedge. Semi-classically, in GR, we can think about $V$ as a particle that starts its trajectory at the past singularity, crosses the past right horizon and falls to the future singularity after passing the future right horizon.

Its companion, $F_V$, however, has an exponentially suppressed tail outside the horizon both in the right {\it and} left wedges (see figure 3).
This follows from the symmetry of the interior filling folded string associated with the condensate $F$,
and hence its bounded excitations.
This is the sense in which the folded string entangles the left and right wedges of the BH and why it should be identified with EPR: the non-perturbative companion of a particle that propagates in the atmosphere of the right wedge has a tail also on the left wedge.

\ifig\loc{{\it Folded strings and EPR}: A particle that lives in the BH atmosphere on the left (a) is accompanied with a non-perturbative mode that lives on the BH-filling folded string (b). The folded string entangles the left and right sides by having tails on both.}
{\epsfxsize4.0in\epsfbox{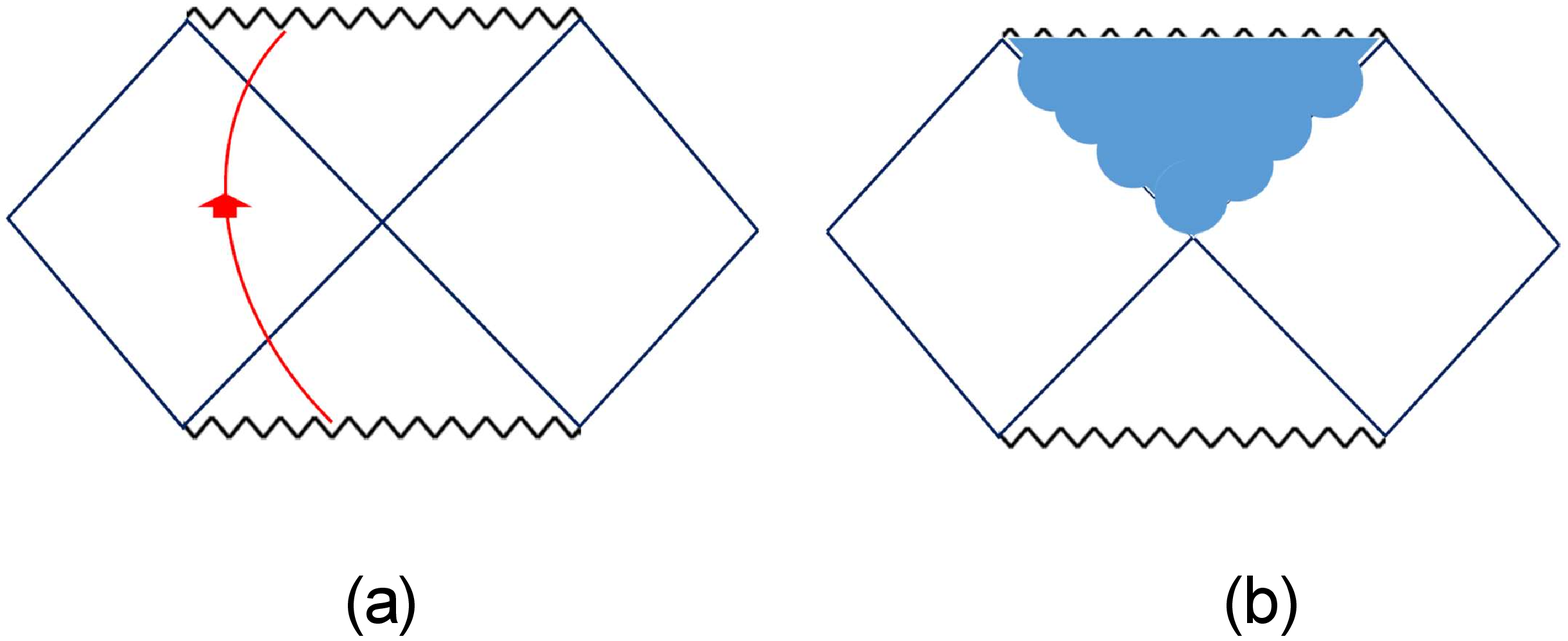}}

While this follows from the $SL(2,\IR)$ structure the question remains: how is this possible? What causes the information that semi-classically
appears only on the right wedge to have (tiny) imprints on the left wedge?
The answer to this was given in \AttaliGOQ. There, the semi-classical energy-momentum tensor associated with the folded string was calculated and it was claimed that if the number of folded strings scales like $1/g_0^2$ then their back-reaction is such that information that falls to the eternal BH, say, from the right side, will not be able to cross the horizon, but rather will get smeared on the future right {\it and} left horizons.  In \GiveonGFK, it was argued that the number of folded strings indeed scales like $1/g_0^2$, 
since $F$ condenses on the sphere. The fact that $F_V$ has a tail also on the left wedge should be viewed as further evidence for these claims.

\newsec{Discussion}

The main point of the paper is to illustrate that in certain cases information in string theory can be quite different
than in quantum field theories.
In particular, each of the GR-like excitations that are bounded to the black fivebrane,
in the sense that they live in the black-fivebrane atmosphere,
is accompanied with a non-perturbative partner that lives on a folded string,
which fills the black-fivebrane interior and has a tiny tail outside the horizon.
The two {\it combined} form the exact state in the theory.
It is the exactness of the underlying $SL(2,\IR)$ CFT that allows us to pinpoint the non-perturbative partner in the case of black fivebranes.
We believe, however, that the existence of a non-perturbative partner, which lives on an extended object in the BH interior,
is likely to be generic in string theory.

We also think that this observation should play a key role in understanding the way information is extracted from black holes.
However, we are still very much confused about the nature of this extraction;
we are not even certain if it is gentle or brutal, in the sense discussed in the introduction.
On the one hand, the identification of modes inside the BH with modes outside the BH
is  in the spirit of the $A=R_B$ and ER=EPR proposals,
which seems to suggest a gentle scenario. On the other hand, the non-perturbative partners do not propagate in an empty BH interior.
Rather, they excite the folded strings that fill the BH interior.
The existence of the  folded strings seems to support a brutal scenario.
In fact, their backreaction prevents information from falling into the BH in a rather non-trivial fashion \AttaliGOQ.
It is as if they were designed for the brutal scenario.

It does feel a bit of a shame to have the key ingredient for the elusive gentle scenario and to end up with a brutal one.
However, right now it is not clear to us how to evade this conclusion.

Perhaps the classical analysis of the folded strings \refs{\ItzhakiGLF,\AttaliGOQ} is misleading,
and it is the appropriate description only at low energies, of the order of the curvature $1/\sqrt{k}$
(that are relevant for the discussion in the previous section)?
Maybe above this energy, we should think about the folded string as a bound state of $k$ gravitons,
in the spirit of $F=I^k$ discussed in subsection 3.5 (continued to the Lorentzian case)?
At large $k$, such a bound state is likely to be close to threshold, from the point of view of an observer at infinity.
In that case, a more energetic mode, with $E\gg 1/\sqrt{k}$,
should be sensitive to the fact that the folded string is made out of a $k$ replica of the ER bridge $I$.

\bigskip
\noindent{\bf Acknowledgements:}
We thank D. Kutasov  for collaboration on some parts of this paper.
This work is supported in part by a center of excellence supported by the Israel Science Foundation (grant number 2289/18)
and BSF (grant number 2018068).

\appendix{A}{$I_{\ell,\bar\ell}$, $W_{\ell,\bar\ell}$ and $F_{\ell,\bar\ell}$}

In this appendix, for completeness,
we present the explicit form of generic $(\ell,\bar\ell)$ excitations,
discussed in section 3,
of the graviton condensate, $I$, the winding string condensate, $W$,
and its fused condensate, $F$, in the worldsheet cigar CFT.
Next, they are listed in turn.

First, from eq. (7.3) in \GiveonDXE, we have
\eqn\ielll{I_{\ell,\bar\ell}=
P_\ell(\partial w,\dots)P_{\bar\ell}(\bar\partial w,\dots)e^{i{p\over\sqrt{2k_b}}x} e^{-\half Q(\ell+\bar\ell)\phi}~,}
where $p$ is the angular momentum on the cigar (see appendix E),
\eqn\ppp{p=\bar\ell-\ell~,}
and from eqs. (7.8) and (7.9) in \GiveonDXE,~\foot{Some sign conventions are different here.} we have
\eqn\welll{W_{\ell,\bar\ell}=e^{i\sqrt{2\over k_b}(mx_L-\bar mx_R)-Q(\tilde j+1)\phi}~,}
where
\eqn\tjmm{\tilde j+1={k_b-\ell-\bar\ell\over 2}~,\qquad (m,\bar m)=\half(k_b+p,k_b-p)~;}
these are winding one operators with momentum $p$ around the cigar (see appendix E).
Finally, the fusion of $W^+\equiv W_{1,1}$ with $W_{\ell,\bar\ell}^*$ is
\eqn\fusionll{F_{\ell,\bar\ell}(w)\sim\int d^2z W^+(z)W_{\ell,\bar\ell}^*(w)~,}
normalized such that, for integer $k$,
\eqn\fffll{F_{\ell,\bar\ell}=P_{k+1-\ell}(\partial w,\dots)P_{k+1-\bar\ell}(\bar\partial w,\dots)
e^{i{p\over\sqrt{2k_b}}x}
e^{-\half Q(2k+2-\ell-\bar\ell)\phi}~.}

\appendix{B}{$\beta^\ell=-P_\ell(\partial w,\dots)$}

In this appendix, we prove eq. \thus\ with \www. One way to do it is the following.
The OPE of the left-moving piece of the winding condensate $W^+(z)$ with its conjugate $W^-(0)$,
defined in eq. \wwww\ with \www, is
\eqn\ope{e^{-w(z)}e^{-\bar w(0)}\sim{1\over z^{k+1}}e^{-[w(z)+\bar w(0)]}
=\sum_n {z^{-1-k+n}\over n!}\left(\partial^n_z e^{-[w(z)+\bar w(0)]}\right)|_{z=0}~,}
hence, for integer $k$, the residue of the pole on the r.h.s is
\eqn\pole{-{1\over k!}P_k(\partial w,\cdots)e^{-{2\over Q}\phi}~,}
where
\eqn\pk{P_k(\partial w,\cdots)=-\left(\partial^k e^{-w}\right)e^w~.}
Now, since the operators $W^\pm$ arise from the left-moving piece of screening operators in the underlying $SL(2,\IR)$ theory,
a.k.a. truly marginal operators which are singlets of the $SL(2,\IR)_L\times SL(2,\IR)_R$ current algebra, the operator \pole\
must arise from a screening operator as well and, since we know the list of screening operators in the $SL(2,\IR)$ theory,
the only candidate is $\beta^k e^{-{2\over Q}\phi}$, reduced to the Euclidean $SL(2,\IR)/U(1)$ coset CFT
and rewritten in the variables of the cigar,
instead of the Wakimoto $\beta-\gamma$ variables in the underlying description (see e.g. \GiveonDXE, for a review).
This is true for any integer $k$, and the normalization is set e.g. knowing that $\beta=\partial w$.
All in all, this leads to \thus\ with \www.

\appendix{C}{Folded string from $AdS_3$}

The main goal of this appendix is to describe the classical folded string in the BH,
from the point of view of $AdS_3$.
The solutions are found via analytic continuation to the winding strings in the cigar.
This gives further support to the stringy HH picture in figure~1.
We also give a geometrical interpretation to \ttt\ and \nowunder.

Lorentzian $AdS_3$ in global coordinates takes the form
\eqn\adsgl{{1\over k}ds^2=-\cosh^2(\rho)d\tau^2+d\rho^2  +\sinh^2(\rho) d\theta^2,}
where $0\leq \rho <\infty,$ $-\infty<\tau<\infty$ and $0<\theta\leq 2\pi.$ This coordinate system is useful when gauging $AdS_3$ to obtain the $SL(2,\IR)/U(1)$ Euclidean cigar, since in that case we merely gauge the $\tau$ direction.

Another coordinate system, which is useful for us, covers the so-called Rindler-AdS space
(see e.g. \ParikhKG, for its definition and geometry).
Lorentzian $AdS_3$ in Rindler coordinates takes the form
\eqn\adsri{{1\over k}ds^2 = -\sinh^2(\rho) dt^2+d\rho^2+\cosh^2(\rho)dx^2,}
where $0\leq \rho <\infty,$ and $-\infty<t,x<\infty$. To obtain the $SL(2,\IR)/U(1)$ BH,
the $x$ direction has to be gauged. Equation \adsri\ describes the region outside the BH.
The horizon is at $\rho=0$. To cross the horizon, we take $\rho\to i\rho$, which gives
\eqn\adstri{{1\over k}ds^2 = \sin^2(\rho) dt^2-d\rho^2+\cos^2(\rho)dx^2.}

To see how \ttt\ and \nowunder\ come about, we note that energy eigenstates behave like
$V_E\sim \exp(-i E t)$, and \MaldacenaHW
\eqn\wads{W^{\pm}\sim \exp(\pm (\theta_L-\theta_R))~,}
where $\theta$ is the angular direction in \adsgl\ and $t$ is the Rindler-AdS time direction.
When $V_E$ goes around $W^{\pm}$ in the worldsheet then, because of \wads, in target space
it means that the wave function associated with $V_E$  makes a full circle, $\theta\to\theta +2\pi$.
Along this circle, it crosses the horizon $4$ times.
Each time the horizon is crossed, $t$ is shifted by $i\beta /4$,
and so by the time the circle is completed, $V_E$ acquires the Boltzmann factor, $\exp(\pm \beta E)$, discussed in section 4.

Next, we turn to the classical folded string solution.
This name is a bit misleading, since as discussed below, in $AdS_3$ the solution is not of a folded string.
Only from the two-dimensional BH perspective, the solution is of a folded string.
To find the solution, we first recall the classical short strings solutions,
found in \MaldacenaHW, that are relevant for the discrete states in the standard $J_3$ basis.
The solutions take the form
\eqn\shst{e^{i\theta}\sinh\rho = i e^{i\omega\tau_1}\sinh\rho_0\,\sin(\alpha \tau_0)~,}
and
\eqn\shstt{\tan\tau={\tan(\omega\tau_0)+\tan(\alpha \tau_0)/\cosh(\rho_0) \over 1- \tan(\omega\tau_0)\tan(\alpha \tau_0)/\cosh(\rho_0)}~,}
where $\tau_0$ and $\tau_1$ are the timelike and spacelike worldsheet coordinates, respectively,
$\omega$ is the winding number around $\theta$,
and $\alpha$ and $\rho_0$ are parameters that describe the solution and are related to the quantum numbers associated with it, \MaldacenaHW.
Since to obtain the $SL(2,\IR)/U(1)$ Euclidean cigar, the $\tau$ direction is gauged away,
the shape associated with this solution in the cigar is determined by \shst\ (and not by \shstt).
It takes the form of a pancake, with a center at the tip of the cigar.
The size of the pancake is basically determined by $\rho_0$.

Note that, while the $AdS_3$ solution, \shst\ and \shstt,
describes \MaldacenaHW\ a winding string that extends in time towards the $AdS_3$ boundary,
till it reaches a maximal size, $\rho_0$, then contracts towards the center,
and repeats to expand and contract periodically in time,
the pancake solution, which is obtained, schematically, by squashing the one in $AdS_3$ in the time direction
(see figure 4),
is thus unoriented in the cigar background.
This means, in particular, that at the classical level, one cannot distinguish between $W^{+}$ and $W^{-}$.
In fact, the pancake solution describes the Euclidean $F$, discussed in subsection 3.3.

\ifig\loc{The pancake solution around the tip of the cigar (on the right) obtained, schematically,
by squashing the short string solution in $AdS_3$ (on the left, \MaldacenaHW).}
{\epsfxsize4.0in\epsfbox{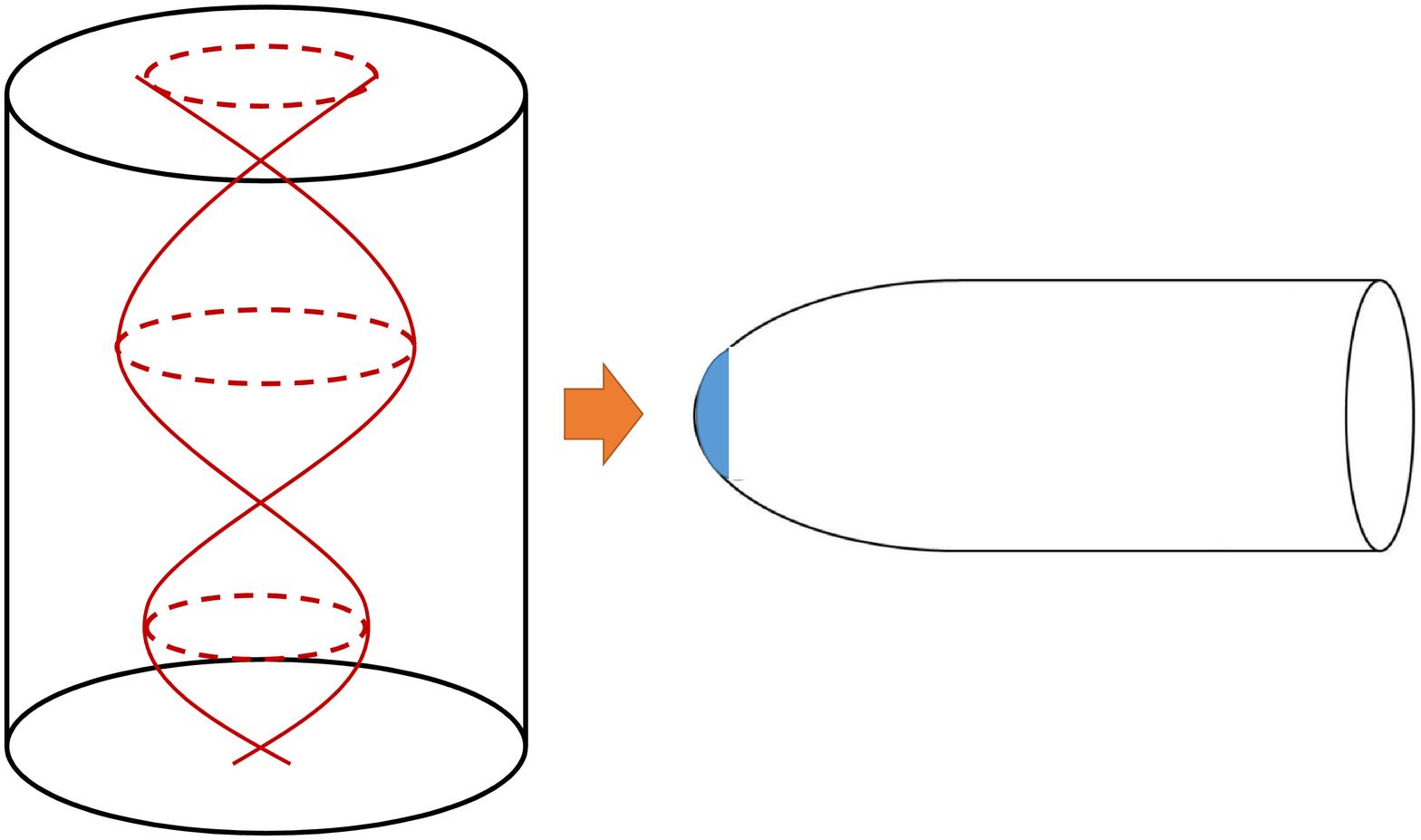}}

We focus on the shortest short string, with $J^a_{L,R}=0$, which corresponds to $W$.
It is obtained by taking $w=\pm 1, \alpha=\mp 1$, and $\rho_0=0$.~\foot{To derive this result, follow subsection 2.3 in \MaldacenaHW.}
The size of the pancake in that case is zero: the pancake collapsed to a point.

To find the folded string solution, we note that  \adsri\ is obtained from \adsgl\ via double analytic continuation,
\eqn\dana{\tau \to i x~,~~~\theta \to i t~.}
Solutions in \adsri, or more precisely in \adstri, can thus be obtained from \shst\ via \dana,  combined with $\rho\to i\rho$.

Starting from \shst\ with $w=-\alpha=1$,
we first shift $\tau_0$ by $\pi/2$, then we double analytically continue on the worldsheet,
\eqn\wsda{\tau_0\to i\sigma_1~,~~~~\tau_1\to i\sigma_0~,}
and in the target space, \dana, to get (after taking $\rho\to i\rho$)
\eqn\fso{t=\sigma_0~,~~~~\sin\rho=\sin \rho_0\,\cosh\sigma_1~,}
where
$-\infty<\sigma_0,\sigma_1<\infty$.

Despite the fact that the range of the new space-like worldsheet coordinate is
$-\infty<\sigma_1<\infty$,
\fso\ describes a closed string.
From the $SL(2,\IR)/U(1)$ BH point of view,
\fso\ is a closed folded string that is created behind the horizon, at $\rho=\rho_0$,
and ends on the singularity; see figure 5.
For $\rho_0\neq 0$, this is not a physical solution in the black fivebrane background,
as it requires a contribution from internal space with a negative dimension.
However, we care about the BH-filling solution with $\rho_0=0$, which is on-shell.
In fact, it is invariant under the $SL(2,\IR)_L \times SL(2,\IR)_R$ affine symmetry,
and so it is natural to identify it with $F$.

\ifig\locc{{\it The solution \fso, reduced on the BH geometry:}
Both $\sigma_0$ and $\sigma_1$ run from $-\infty$ to $\infty$.
They do so in a way that at $\rho=\rho_0$ (at the bottom of the shaded regime)
and $t=i\beta/4$ (namely, at $t=0$ in the BH interior),
a closed folded string is created, propagates (in global time, namely, upwards) and ends at the singularity.
The BH-filling string amounts to the $\rho_0\to 0$ limit.
}
{\epsfxsize4.0in\epsfbox{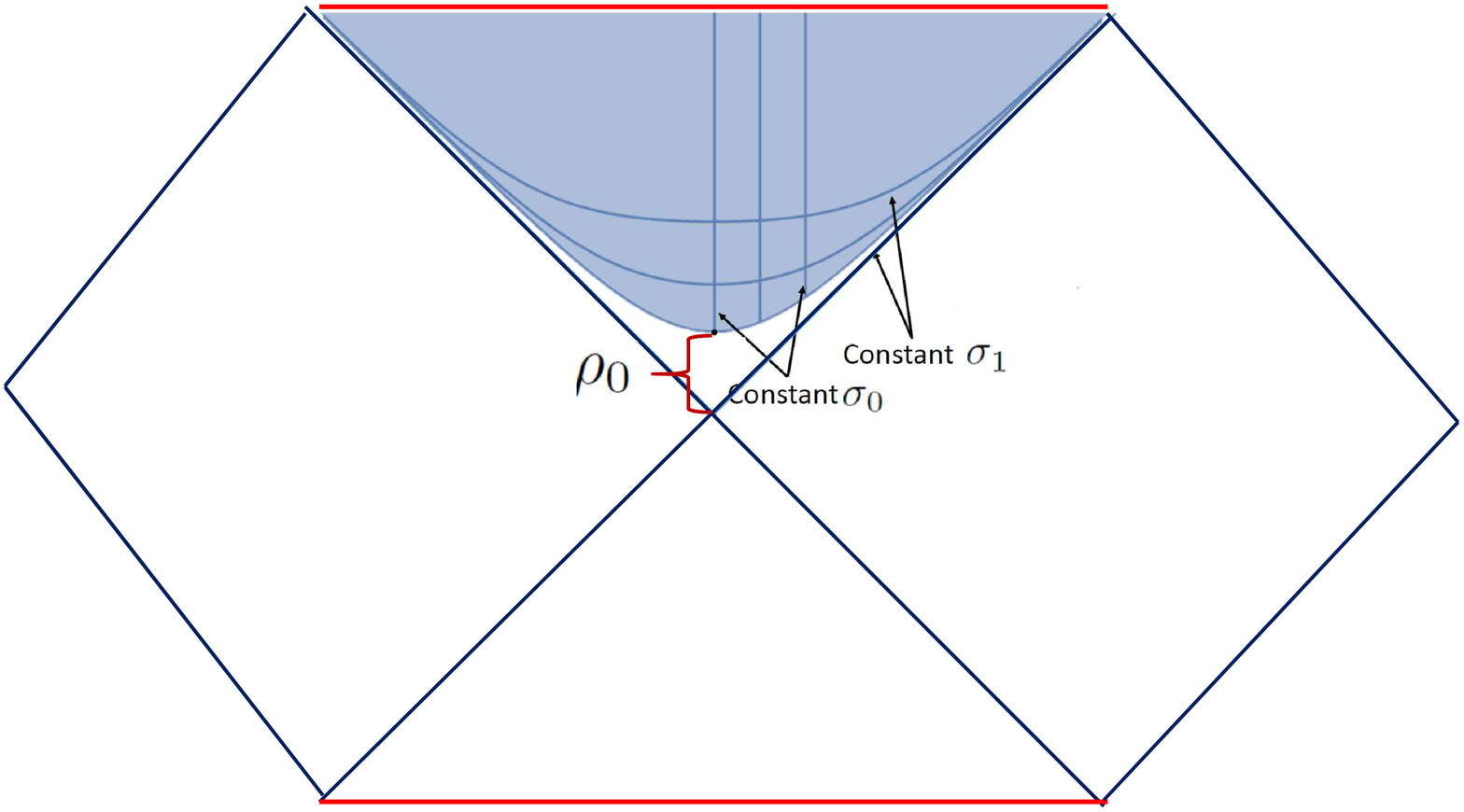}}

The reasoning of \refs{\MaldacenaHI,\ItzhakiGLF} implies that there are many other on-shell solutions in the two-dimensional BH geometry.
These, however, do not have specific quantum numbers under the $SL(2,\IR)_L \times SL(2,\IR)_R$,
and so cannot be obtained from \shst\ and \shstt.

From the $AdS_3$ point of view, the semiclassical configuration in eq. \fso, unlike the one in eq. \shst, ends on the boundary,
and so in the dual space-time $CFT_2$ it does not correspond to a local operator.
This is yet another aspect why $W$ is more fundamental than the $F$ condensate.

\appendix{D}{The construction of $W_V$}

To obtain the GFZZ duals, $W_V$ in \vvwwje, of the GR bound states, $V$ in \vje,
it is convenient to write the non-compact, space-like $J_2$ eigenstates
in the underlying $SL(2,\IR)$ (the hyperpolic basis; see e.g. section 4 of \LindbladNigel)
in terms of the standard basis of the compact, time-like $J^3$ eigenstates,
\eqn\jmjm{V_{j;m_2}=\sum_{m_3} A_{m_3}(j,m_2)V_{j;m_3}~,}
where here we restrict to the left-moving piece of $V_{jE}$, for simplicity,
whose parent in $SL(2,\IR)$ we denote by $V_{j;m_2}$.
For states in the $D^+_j$ principal discrete representations (see e.g. \MaldacenaHW\ for definitions and details), \jmjm\ takes the explicit form
\eqn\ejmjm{V_{j;m_2}=\sum_{m_3=j+1}^\infty A_{m_3}(j,m_2)(J_0^+)^{m_3-j-1} V_{j;j+1}~.}
Now, via the isomorphism between the principal discrete representation, $\hat D^{+,\omega=0}_j$, in affine $SL(2,\IR)$,
and its spectrally-flowed (a.k.a. twisted) $\hat D^{-,\omega=1}_{\tilde j={k\over 2}-j-1}$ representation \MaldacenaHW,~\foot{We
consider again the bosonic case here, for simplicity;
the extension to the fermionic case is straightforward \refs{\GiveonDXE,\unpublished}.}
we know that the operator
\eqn\vjwm{\Phi_{\tilde j;m_2}^{-,\omega=1}=\sum_{N=0}^\infty A_{j+1+N}(j,m_2)(J_{-1}^+)^N\Phi^{\omega=1}_{\tilde j;-\tilde j-1}~,
\qquad\tilde j={k\over 2}-j-1~,}
where $\Phi^\omega$ denotes a spectrally flowed, a.k.a. a twisted operator, in the $\omega$ sector
(see \MaldacenaHW\ and \ArgurioTB\ for details and notation),
is related to \ejmjm\ by GFZZ duality \GiveonDXE, and
the two create the {\it same} state.

Finally, the reduction of these $J^2$ eigenstates to the Lorentzian $SL(2,\IR)/U(1)$ coset, $V^{\omega=1}_{\tilde jE}$,
is obtained by decomposing
\eqn\decomp{\Phi_{\tilde j;m_2}^{-,\omega=1}=V^{\omega=1}_{\tilde jE}e^{iEx_2}~,}
where the energy $E$ is related to the $J^2$ eigenvalue $m_2$ by eq. \eqm,
and $x_2$ is a canonically normalized scalar bosonizing the $J^2$ current,
\eqn\jtwo{J^2={1\over Q}i\partial x_2~.}
Now, together with the ghosts and internal pieces, they give rise to the GFZZ duals, $W_V$,
of the bound states $V$ in \vje,
\eqn\vwje{W_V^+=e^{-\varphi-\bar\varphi}V^{\omega=1}_{\tilde jE}V_\NN~.}
The conjugates, $W_V^-$, are obtained, similarly, starting with the $\hat D^-_j$ principal discrete representations in \ejmjm.

A few comments on the properties of \vjwm,\vwje\ are in order:
\item{*}
The energy eigenstates in the two-dimensional BH, which are reduced from their $SL(2,\IR)$ parent, \vjwm,
consist of an infinite sum over string excitations,
the $N$'th excitation contributing with its corresponding coefficient, $A_{j+1+N}(j,m_2)$.
\item{*}
These $A_{m_3}(j,m_2)\equiv\langle j,m_3|j,m_2\rangle$, in eqs. \jmjm--\vjwm, can be found in eq. (4.17) of \LindbladNigel,
where the study of continuous bases for unitary irreducible representations of $SL(2,\IR)$ is presented;
note, in particular, that $\langle j,m_2|j,m_2'\rangle=\delta(m_2-m_2')$, as expected for bound states in the BH zone.
\item{*}
The wave functions of the stringy contribution to the bound states -- the $W_V$ in eq. \vwje,
behave like $e^{-({1\over Q}-Qj)\phi}$, at large radial $\phi$, as do the excitations of
the winding string condensate in the Euclidean case -- the $W_\ell$ in eq. \well,
due to the $\Phi^{\omega=1}_{\tilde j={k\over 2}-j-1}$ piece in eq. \vjwm,
and thus they are highly localized near the horizon, when the curvature and imaginary radial momentum are small, $Q,j/k\ll 1$,
as opposed to their GFZZ duals, GR bound states -- the $V$ in eq. \vje,
whose wave functions behave like $e^{-Q(j+1)\phi}$, instead, as do the excitations of the graviton condensate in the Euclidean case
-- the $I_\ell$ in eq. \iell, and are thus spread over the BH atmosphere.

\appendix{E}{Some facts on $SL(2,\IR)/U(1)$}

In this appendix, we collect some facts concerning properties of the $SL(2,\IR)/U(1)$ quotient CFTs
that are used in the text. The reader is referred to e.g. \GiveonDXE\ for more facts and details.

Operators $V_j$ in $SL(2,\IR)/U(1)$ quotient CFTs descend from operators $\Phi_j$ in the underlying $SL(2,\IR)$ theory,
and carry, in particular, the quantum number $j$ of the $SL(2,\IR)$ representation, as well as other quantum numbers,
depending on the particular choice of basis and  subjected to the gauge condition imposed by the Abelian gauging done.
The quantum number $j$ governs the radial dependence of the wave functions of states in the resulting sigma-model geometry.
Concretely, the vertex operator $V_j$ decays at large $\phi$ as
\eqn\vjphi{V_j\sim e^{-Q(j+1)\phi}~.}
Stripping off the factor of the string coupling $e^{-Q\phi/2}$, \lindil,
that relates the vertex operator to the wave function, we find that the wave function of the state behaves at large $\phi$
as $e^{-Q(j+1/2)\phi}$, and hence it is normalizable if $j>-1/2$.
However, unitarity of the CFT leads to a more restrictive bound on $j\in\IR$ -- the unitarity bound:
\eqn\unitarity{-\half<j<{k-1\over 2}~,}
where $k$ is the total level of the $SL(2,\IR)$ theory,
\eqn\kkbq{k=k_b-2=\sqrt{2\over Q}~,}
and $k_b$ is the bosonic level.

For instance, the cigar CFT is obtained by an axial gauging in the compact time-like direction  of $SL(2,\IR)$,
and a convenient basis of states is thus eigenstates of the generators $(J^3,\bar J^3)$, whose eigenvalues are denoted by $(m,\bar m)$.
In the coset CFT, the latter are subject to the gauge condition
\eqn\plpr{\sqrt{2\over k_b}(m,-\bar m)=(p_L,p_R)\equiv\left({p\over R}+{\omega R\over 2},{p\over R}-{\omega R\over 2}\right), \qquad R=\sqrt{2k_b},
\qquad p,\omega\in Z;}
the integers $p$ and $w$ are thus the quantized momentum and winding, respectively, on the asymptotic cylinder of the cigar.

A large class of states in the cigar CFT is described by Virasoro primary vertex operators, $V_{j;m,\bar m}$.
Far from the tip of the cigar, they behave as
\eqn\vjmfar{V_{j;m,\bar m}\simeq e^{ip_Lx_L+ip_Rx_R-Q(j+1)\phi}~,}
where $(p_L,p_R)$ is given in \plpr.
The operators $W_{\ell,\bar\ell}$ in section 3 and appendix A are examples of such operators,
with $\omega=1$ and $p=\bar\ell-\ell$.
The operators $I_{\ell,\bar\ell}$ and $F_{\ell,\bar\ell}$ have non-zero oscillations modes
on the cigar, the $P_\ell(\partial w,\dots)$ and $P_{\bar\ell}(\bar\partial w,\dots)$ in section 3 and appendices A and B,
on top of an \vjmfar\ piece with   $\omega=0$.

\listrefs

\end